\documentclass[prb,superscriptaddress,showpacs,floatfix,tightenlines,10pt,twocolumn]{revtex4-1}
\usepackage{amssymb}
\usepackage{amsmath}
\usepackage{graphicx}
\usepackage{float}
\usepackage{xcolor}

\newcommand{\func}[1]{\operatorname{#1}}

\begin{document}

\title{Magneto-optical Kerr effect in Weyl semimetals with broken inversion
and time-reversal symmetries}
\author{Olivier Tr\'{e}panier}
\author{R\'{e}mi N. Duchesne}
\author{J\'{e}r\'{e}mie J. Boudreault}
\author{Ren\'{e} C\^{o}t\'{e}}
\affiliation{D\'{e}partement de physique and Institut Quantique, Universit\'{e} de
Sherbrooke, Sherbrooke, Qu\'{e}bec, Canada J1K 2R1 }
\date{\today }

\begin{abstract}
The topological nature of the band structure of a Weyl semimetal leads to a
number of unique transport and optical properties. For example, the
description of the propagation of an electromagnetic wave in a Weyl
semimetal with broken time-reversal and inversion symmetry, for example,
requires a modification of the Maxwell equations by the axion field $\theta
\left( \mathbf{r},t\right) =2\mathbf{b}\cdot \mathbf{r}-2b_{0}t,$ where $2%
\mathbf{b}$ is the separation in wave vector space between two Weyl nodes of
opposite chiralities and $2\hslash b_{0}$ is their separation in energy. In
this paper, we study theoretically how the axion terms $b_{0}$ and $\mathbf{b%
}$ modify the frequency behavior of the Kerr rotation and ellipticity angles 
$\theta _{K}\left( \omega \right) $ and $\psi _{K}\left( \omega \right) $ in
a Weyl semimetal. Both the Faraday and Voigt configurations are considered
since they provide different information on the electronic transitions and
plasmon excitation. We derive the Kerr angles firstly without an external
magnetic field where the rotation of the polarization is only due to the
axion terms and secondly in a strong magnetic field where these terms
compete with the gyration effect of the magnetic field. In this latter case,
we concentrate on the ultra-quantum limit where the Fermi level lies in the
chiral Landau level and the Kerr and ellipticity angles have more complex
frequency and magnetic field behaviors.
\end{abstract}

\maketitle

\section{INTRODUCTION}

A Weyl semimetal\cite{Review} (WSM) is a three-dimensional topological phase
of matter where pairs of nondegenerate bands cross at isolated points in the
Brillouin zone called "Weyl nodes". Near these points, the electronic
dispersion is gapless and linear in wave vector. The electrons in each node
are described by the Weyl equation which is a two-component analog of the
Dirac equation and to each node is associated a chirality index $\chi =\pm 1$
which reflects the topological nature of the band structure. The
Nielsen-Ninomiya theorem\cite{Nielsen} requires that the number of Weyl
points in the Brillouin zone be even so that Weyl nodes must occur in pairs
of opposite chiralities. Stability of the Weyl nodes requires that
time-reversal and/or inversion symmetry be broken. If both symmetries are
broken, the minimal number of Weyl nodes is two.

The topological nature of the band structure of a WSM leads to a number of
interesting transport and optical properties such as an anomalous Hall effect%
\cite{AHE}, the chiral-magnetic effect\cite{CME}, Fermi arcs\cite{FermiArc},
a chiral anomaly leading to a negative longitudinal negative
magnetoresistance\cite{ChiralAnomaly} and to a plasmon\cite{Spivak} whose
frequency increases with the magnetic field $B$ as $\sqrt{B}$.

Valuable information on the band structure and topological properties of
WSMs can be obtained by magneto-optical methods such as the magneto-optical
Kerr effect (MOKE)\cite{Cheskis} which is a powerful probe of the magnetic
and electronic properties of a material. The MOKE\ is a noncontact technique
that measures the rotation of the polarization vector and the ellipticity of
an electromagnetic wave reflected from the surface of a material. The
rotation is due to the presence of off-diagonal elements in the dielectric
tensor of a material that are caused by an external magnetic field and, in
WSMs, by the axion terms $b_{0}$ and $\mathbf{b}$ in addition to the
magnetic field. The transmission analog of the MOKE, the Faraday effect, has
been shown to be very large in some WSMs\cite{Han2022} and, as we
demonstrate in this paper, it is also the case for the Kerr effect.

In order to study the optical properties of WSMs, one must take into account
the fact that the source terms in Maxwell equations are modified by the
axion field $\theta \left( \mathbf{r},t\right) =2\mathbf{b}\cdot \mathbf{r}%
-2b_{0}t,$ where $2\hslash \mathbf{b}$ in the separation in momentum space
between two Weyl nodes of opposite chiralities and $2\hslash b_{0}$ is their
separation in energy\cite{Wilczek,Wu, Burkov2012}. The axion terms generate
an extra current $\mathbf{j}=-\frac{e^{2}}{2\pi ^{2}\hslash }\mathbf{b}%
\times \mathbf{E}+\frac{e^{2}}{2\pi ^{2}\hslash }b_{0}\mathbf{B}$ in the Amp%
\`{e}re-Maxwell equation and an extra density $\rho =\frac{2\alpha }{\pi }%
\sqrt{\frac{\varepsilon _{0}}{\mu _{0}}}\mathbf{b}\cdot \mathbf{B}$ in the
Gauss law (here, $\alpha =e^{2}/4\pi \varepsilon _{0}\hslash c$ is the
fine-structure constant and $\varepsilon _{0},\mu _{0}$ are the permittivity
and permeability of free space). The first axion term, $\mathbf{b,}$ is
present if time-reversal symmetry is broken and is responsible for the
anomalous Hall effect while the second axion term, $b_{0},$ breaks
space-inversion symmetry and leads to the chiral magnetic effect. Even in
the absence of a magnetic field, a WSM has a gyrotropic property if $\mathbf{%
b}$ or (to a lesser extent) $b_{0}$ is non-zero (as we will show, the
rotating power of $b_{0}$ is negligible in comparison with that of $\mathbf{b%
}$)$\mathbf{.}$

The effect of the axion term $\mathbf{b}$ on the Kerr and Faraday rotations
has been studied previously\cite{Kargarian} using a simple, undoped,
two-node model of a WSM with broken time-reversal symmetry and in zero
magnetic field. This work has subsequently been extended to include doping
and tilting of the Weyl cones in Ref. \onlinecite{Sonowal2019} where the
authors predicted a giant Kerr angle of the order of $0.1$ rad for optical
frequencies below $10$ THz in both type-I and type-II\ thin films of Weyl
semimetals. They also found a large Kerr effect caused by the axion term $%
\mathbf{b}$ in a bulk WSM. In a previous paper by one of us\cite{Parent2020}%
, it was shown that the MOKE\ in a Weyl semimetal with time-reversal
symmetry (i.e., $\mathbf{b}=0$) and tilted Weyl cones in a strong magnetic
field is modified in comparison with its behavior in a normal metal. It was
also shown that a valley polarization\cite{Bertrand2019} can be detected in
the MOKE\ signal which also bears a signature of the chiral anomaly in the
form of a large peak at the plasmon frequency in the Kerr angle in the Voigt
configuration. This peak is blueshifted as $\omega \sim \sqrt{B}$ as expected%
\cite{Spivak}.

In this paper, we study the frequency and magnetic field behaviors of the
Kerr and ellipticity angles $\theta _{K}\left( \omega ,B\right) $ and $\psi
_{K}\left( \omega ,B\right) $ in a WSM with broken time-reversal and
inversion symmetries with: 1) the axion terms $\mathbf{b}$ and $b_{0}$ alone 
\textit{(}i.e., the natural gyrotropy of the WSM) and 2) in the presence of
a strong magnetic field where these terms compete with the gyrotropy that
the magnetic field induces. We work at $T=0$ K with a two-node model, the
simplest possible in order to get the essential physics. We allow for doping
so that, in conjunction with $b_{0}$, the WSM can be electron or hole doped
or at compensation. The Dirac cones are untilted and the WSM\ is in
equilibrium with a common Fermi level $e_{F}$ for both nodes.

In a strong magnetic field, the electronic energy is quantized into a set of
negative and positive energy levels $n=0,\pm 1,\pm 2,...$. An important
characteristic of a WSM is the presence of a single linearly dispersive
chiral Landau level (the $n=0$ level) at each node where the electrons
disperse in only one direction which is dictated by the chirality index $%
\chi $ of the node. This chiral Landau level modifies the dielectric tensor
and so the optical properties of a WSM. This is especially true in the
ultra-quantum regime where the Fermi level $e_{F}$ lies in this chiral
level. The lowest-energy electronic transitions are then given by the
frequencies $\omega _{n,m}$ with $n=-1,m=0$ and $n=0,m=1.$ In this regime,
the Kerr angle is particularly interesting. Indeed, we find a plateau where $%
\theta _{K}\left( \omega \right) =\pi /2$ and another one where $\theta
_{K}\left( \omega \right) =0$ when $\omega $ is smaller than the value $\min
\left( \omega _{-1,0},\omega _{0,1}\right) $ i.e., in the Pauli blocked
regime. We also find that for the special case where $b\hslash v_{F}=e_{F}$
for $\mathbf{b}\Vert \mathbf{B}$ and $b_{0}=0$ ($v_{F}$ is the Dirac-Weyl
velocity), the Kerr angle is exactly zero for $\omega \in \left[ 0,\min
\left( \omega _{-1,0},\omega _{0,1}\right) \right] .$

In a magnetic field, two configurations are usually studied\cite{Balkanski}.
One is the Faraday configuration where the incident light propagates along
the direction of the magnetic field $\mathbf{B}$ with its polarization
vector perpendicular to $\mathbf{B}$ and the Voigt configuration where the
propagation is perpendicular to $\mathbf{B}$ but the polarization vector has
a component along $\mathbf{B}.$ The two configurations give different
information on the system capturing, for example, different electronic
transitions. In a WSM, we find that the plasmon mode has a signature in $%
\theta _{K}\left( \omega \right) $ in the Voigt configuration only if $%
\mathbf{b}\Vert $ $\mathbf{B}$ while it is signaled in both configurations
if $\mathbf{b}$ is tilted with respect to $\mathbf{B}.$

The remainder of this paper is organized as follows:\ we derive the
dielectric tensor of a WSM with and without magnetic field in Sections II\
and III. Section IV discusses the modification of the Maxwell equations by
the axion terms. We introduce the formalism necessary to compute the Kerr
and ellipticity angles in the Faraday and Voigt configurations in Sec. V. We
briefly review the Kerr rotation for a normal metal in Sec. VI and then
present our numerical results for the frequency dependence of the Kerr angle
in a WSM\ with and without magnetic field in Secs. VII and VIII. We study
the effect of tilting the vector $\mathbf{b}$ with respect to the magnetic
field in Sec. IX. In Section X, we discuss the magnetic field and $\mathbf{b}
$ dependence (for $\mathbf{b}\Vert \mathbf{B}$) of the Kerr angle in the
Faraday configuration. We conclude in Sec. XI.

\section{DIELECTRIC TENSOR OF A WSM\ IN ZERO MAGNETIC FIELD}

We consider a simple two-node model of a Weyl semimetal (WSM) with broken
space-inversion and time-reversal symmetries so that the two nodes have
opposite chiralities. The nodes are centered at wave vectors $-\chi \mathbf{b%
}$, where $\chi =\pm 1$ is the chirality index. The Hamiltonian of the
electron gas in each Weyl node is given by 
\begin{equation}
h_{\chi }\left( \mathbf{k}\right) =\chi \hslash v_{F}\mathbf{k}\cdot \mathbf{%
\sigma }+\chi \hslash b_{0}I_{2},  \label{hamil}
\end{equation}%
where $v_{F}$ is the Dirac-Weyl velocity, $\mathbf{k}$ is the wave vector
measured from the position of the Weyl node, $\mathbf{\sigma }$ is the
vector of Pauli matrices defined in the basis of the two electronic bands
that cross, $I_{2}$ is the $2\times 2$ unit matrix, $2\hslash b_{0}$ is the
separation in energy of the two Dirac points, and $2\hslash \mathbf{b}$ is
their separation in momentum space. The former term breaks the
space-inversion symmetry while the latter breaks time-reversal symmetry.

The energy spectrum is given by

\begin{equation}
E_{\chi ,s}\left( \mathbf{k}\right) =s\hslash v_{F}\left\vert \mathbf{k}%
\right\vert +\chi \hslash b_{0},
\end{equation}%
where $s=\pm 1$ is the band index i.e., the upper ($s=+$) or lower part ($%
s=- $) of the Dirac cone. The corresponding eigenvectors are of the form $%
v_{\chi ,\mathbf{k},s}\left( \mathbf{r}\right) =\frac{1}{\sqrt{V}}\eta
_{\chi ,s}\left( \mathbf{k}\right) e^{i\mathbf{k}\cdot \mathbf{r}}$, where $%
V $ is the volume of the WSM and the spinors $\eta _{\chi ,s}\left( \mathbf{k%
}\right) $ are given by%
\begin{eqnarray}
\eta _{+,+}\left( \mathbf{k}\right) &=&\eta _{-,-}\left( \mathbf{k}\right)
=\left( 
\begin{array}{c}
e^{-i\varphi _{\mathbf{k}}/2}\cos \left( \frac{\theta _{\mathbf{k}}}{2}%
\right) \\ 
e^{i\varphi _{\mathbf{k}}/2}\sin \left( \frac{\theta _{\mathbf{k}}}{2}\right)%
\end{array}%
\right) , \\
\eta _{+,-}\left( \mathbf{k}\right) &=&\eta _{-,+}\left( \mathbf{k}\right)
=\left( 
\begin{array}{c}
-e^{-i\varphi _{\mathbf{k}}/2}\sin \left( \frac{\theta _{\mathbf{k}}}{2}%
\right) \\ 
e^{i\varphi _{\mathbf{k}}/2}\cos \left( \frac{\theta _{\mathbf{k}}}{2}\right)%
\end{array}%
\right) ,
\end{eqnarray}%
where $\theta _{\mathbf{k}},\varphi _{\mathbf{k}}$ are respectively the
polar and azimuthal angles of the wave vector $\mathbf{k}.$

To derive the many-body Hamiltonian, we define the field operators%
\begin{equation}
\Psi _{\chi }\left( \mathbf{r}\right) =\frac{1}{\sqrt{V}}\sum_{s,\mathbf{k}%
}e^{i\mathbf{k}\cdot \mathbf{r}}\eta _{\chi ,s}\left( \mathbf{k}\right)
c_{\chi ,s,\mathbf{k}}.  \label{field}
\end{equation}%
The operators $c_{\chi ,s,\mathbf{k}},c_{\chi ,s,\mathbf{k}}^{\dag }$
annihilate or create an electron in state $\left( \chi ,s,\mathbf{k}\right)
. $They satisfy the commutation relation $\left\{ c_{\chi ,s,\mathbf{k}%
},c_{\chi ^{\prime },s^{\prime },\mathbf{k}^{\prime }}^{\dag }\right\}
=\delta _{\chi ,\chi ^{\prime }}\delta _{s,s^{\prime }}\delta _{\mathbf{k},%
\mathbf{k}^{\prime }}.$ The second-quantized Hamiltonian for each node is
then%
\begin{eqnarray}
H_{\chi } &=&\int d^{3}r\Psi _{\chi }^{\dag }\left( \mathbf{r}\right)
h_{\chi }\left( \mathbf{k}\right) \Psi _{\chi }\left( \mathbf{r}\right) \\
&=&\sum_{s,\mathbf{k}}E_{\chi ,s}\left( \mathbf{k}\right) c_{\chi ,s,\mathbf{%
k}}^{\dag }c_{\chi ,s,\mathbf{k}}.  \notag
\end{eqnarray}

The current operator is obtained, after making the Peierls substitution $%
h_{\chi }\left( \mathbf{k}\right) \rightarrow h_{\chi }\left( \mathbf{k}+e%
\mathbf{A}/\hslash \right) $ in the hamiltonian ($e>0$ for an electron and $%
\mathbf{A}_{e}$ is the vector potential of an external electromagnetic
field), by taking the derivative 
\begin{equation}
\mathbf{j}_{\chi }=-\frac{\delta h_{\chi }}{\delta \mathbf{A}_{e}}=-\chi
ev_{F}\mathbf{\sigma ,}  \label{courant}
\end{equation}%
so that the total current at each node is given by 
\begin{eqnarray}
\mathbf{J}_{\chi } &=&\int d^{3}r\Psi _{\chi }^{\dag }\left( \mathbf{r}%
\right) \mathbf{j}_{\chi }\Psi _{\chi }\left( \mathbf{r}\right)
\label{current} \\
&=&-e\chi v_{F}\sum_{s,s^{\prime },\mathbf{k}}\left[ \eta _{\chi ,s}^{\dag
}\left( \mathbf{k}\right) \mathbf{\sigma }\eta _{\chi ,s^{\prime }}\left( 
\mathbf{k}\right) \right] c_{\chi ,s,\mathbf{k}}^{\dag }c_{\chi ,s^{\prime },%
\mathbf{k}}.  \notag
\end{eqnarray}%
The current operator enters the definition of the retarded current-current
response tensor $\overleftrightarrow{\mathbf{\chi }}_{\chi }^{R}\left(
\omega \right) $ which is obtained by taking the analytic continuation $%
i\Omega _{n}\rightarrow \omega +i\delta $ of the two-particle Matsubara
Green's function%
\begin{equation}
\overleftrightarrow{\mathbf{\chi }}_{\chi }\left( i\Omega _{n}\right) =-%
\frac{1}{V\hslash }\int_{0}^{\beta \hslash }d\tau \left\langle T\mathbf{J}%
_{\chi }\left( \tau \right) \mathbf{J}_{\chi }\left( 0\right) \right\rangle
e^{i\Omega _{n}\tau },  \label{deux}
\end{equation}%
where $\tau $ is the imaginary time, $\Omega _{n}$ is a bosonic Matsubara
frequency and $\beta =1/k_{B}T$ with $k_{B}$ the Boltzmann constant. The
optical conductivity tensor can then be calculated from the expression%
\begin{equation}
\overleftrightarrow{\mathbf{\sigma }}\left( \omega \right) =\frac{i}{\omega
+i\delta }\sum_{\chi }\left[ \overleftrightarrow{\mathbf{\chi }}_{\chi
}\left( \omega \right) -\overleftrightarrow{\mathbf{\chi }}_{\chi }\left(
0\right) \right] .  \label{un}
\end{equation}%
In the absence of an external static magnetic field and for the simple
Hamiltonian given in Eq. (\ref{hamil}), $\sigma _{i,j}\left( \omega \right)
=\delta _{i,j}\sigma \left( \omega \right) $ so that we only need to compute
the response function 
\begin{eqnarray}
\chi _{\chi }\left( \omega \right) &=&-\frac{e^{2}v_{F}^{2}}{\hslash V}%
\sum_{s,s^{\prime },\mathbf{k}}\left\vert \eta _{\chi ,s}^{\dag }\left( 
\mathbf{k}\right) \sigma ^{\left( \alpha \right) }\eta _{\chi ,s^{\prime
}}\left( \mathbf{k}\right) \right\vert ^{2}  \label{chic} \\
&&\times \frac{\left\langle n_{\chi ,s^{\prime }}\left( \mathbf{k}\right)
\right\rangle -\left\langle n_{\chi ,s}\left( \mathbf{k}\right)
\right\rangle }{\omega +i\delta -\left( E_{\chi ,s^{\prime }}\left( \mathbf{k%
}\right) -E_{\chi ,s}\left( \mathbf{k}\right) \right) /\hslash },  \notag
\end{eqnarray}%
where $\left\langle n_{\chi ,s}\left( \mathbf{k}\right) \right\rangle
=\left\langle c_{\chi ,s,\mathbf{k}}^{\dag }c_{\chi ,s,\mathbf{k}%
}\right\rangle $ is the average occupation of the state $\left( \chi ,s,%
\mathbf{k}\right) $ at zero temperature and $\alpha $ can be any of $x,y,z.$

The relative dielectric tensor is 
\begin{equation}
\varepsilon \left( \omega \right) =\varepsilon _{b}+\frac{i\sigma \left(
\omega \right) }{\varepsilon _{0}\left( \omega +i\delta \right) },
\label{trois}
\end{equation}%
where $\varepsilon _{b}$ is the bound charge contribution. We take $%
\varepsilon _{b}=1$ in our calculations, but comment later on the effect of
a larger $\varepsilon _{b}$ on the Kerr rotation. We assume that the two
nodes are at a thermodynamical equilibrium with a Fermi level $e_{F}$ so
that the Fermi wave vector $k_{F,\chi }$ and the Fermi level $e_{F,\chi }$
for each node are related by 
\begin{equation}
e_{F,\chi }=\hslash v_{F}k_{F,\chi }=e_{F}-\chi \hslash b_{0}.
\label{quatre}
\end{equation}%
The optical conductivity has contributions from the interband and intraband
transitions. Both contributions have been calculated in parts in a number of
papers\cite%
{Kargarian,Sonowal2019,Carbotte2014,Carbotte2016,Carbotte2018,Carbotte2021}.
For completeness, we derive the intraband contribution in Appendix A. In the
absence of magnetic field and in the continuum model, the relative
dielectric function is

\begin{eqnarray}
\varepsilon \left( \omega \right) &=&1  \label{epsiwsm} \\
&&+\gamma \sum_{\chi }\left[ i\theta \left( \omega -2v_{F}k_{F,\chi }\right)
+\frac{1}{\pi }\ln \left( \left\vert \frac{\omega ^{2}-4v_{F}^{2}k_{c}^{2}}{%
\omega ^{2}-4v_{F}^{2}k_{F,\chi }^{2}}\right\vert \right) \right]  \notag \\
&&+\gamma \sum_{\chi }\frac{4}{\pi \hslash ^{2}\omega }\frac{i\tau }{%
1+\omega ^{2}\tau ^{2}}\left( e_{F,\chi }^{2}+\frac{1}{3}\hslash ^{2}\omega
^{2}+\frac{\hslash ^{2}}{4\tau ^{2}}\right)  \notag \\
&&-\gamma \sum_{\chi }\frac{4}{\pi \hslash ^{2}}\frac{\tau ^{2}}{1+\omega
^{2}\tau ^{2}}\left( e_{F,\chi }^{2}-\frac{\hslash ^{2}}{12\tau ^{2}}\right)
,  \notag
\end{eqnarray}%
where $\gamma =\frac{\alpha }{6v_{F}/c},k_{c}$ is a cutoff wave vector which
is related to the applicability of the linear dispersion relation and $\tau $
is the scattering time that enters in the calculation of the intraband
conductivity (see Appendix A). The last two lines in Eq. (\ref{epsiwsm}) are
intraband contributions to the dielectric function.

\section{DIELECTRIC TENSOR OF A\ WSM\ IN A MAGNETIC FIELD}

In a magnetic field $\mathbf{B}=\mathbf{\nabla }\times \mathbf{A=}B\widehat{%
\mathbf{z}}$ and in the Landau gauge $\mathbf{A}=\left( 0,Bx,0\right) $, the
single-particle Hamiltonian becomes, after making the Peierls substitution $%
\mathbf{k\rightarrow k}+e\mathbf{A}/\hslash ,$ 
\begin{equation}
h_{\chi }\left( \mathbf{k}\right) =\chi \hslash b_{0}I_{2}+\chi \hslash
v_{F}\left( 
\begin{array}{cc}
k_{z} & \frac{\sqrt{2}}{\ell }a \\ 
\frac{\sqrt{2}}{\ell }a^{\dag } & -k_{z}%
\end{array}%
\right) ,  \label{h5}
\end{equation}%
where $I_{2}$ is the $2\times 2$ unit matrix and the ladder operators are
defined by%
\begin{eqnarray}
a &=&\frac{\ell }{\sqrt{2}\hslash }\left( P_{x}-iP_{y}\right) , \\
a^{\dag } &=&\frac{\ell }{\sqrt{2}\hslash }\left( P_{x}+iP_{y}\right) ,
\end{eqnarray}%
where $\mathbf{P}\equiv \hslash \mathbf{k}+e\mathbf{A.}$They obey the
commutation relation $\left[ a,a^{\dag }\right] =1.$ Here $h_{\chi }\left( 
\mathbf{k}\right) $ is written in the same basis as that used in Eq. (\ref%
{hamil}).

The energy spectrum of a Weyl node now consists of a \textit{chiral} Landau
level (index $n=0$) which disperses in one direction only according to%
\begin{equation}
E_{\chi ,0}\left( k\right) =\chi \frac{\hslash v_{F}}{\ell }\left( -k\ell +%
\frac{b_{0}\ell }{v_{F}}\right)
\end{equation}%
and of a set of positive and negative energy Landau levels $n=1,2,3,...$
with dispersion%
\begin{equation}
E_{\chi ,n>0,s}\left( k\right) =\frac{\hslash v_{F}}{\ell }\left( s\sqrt{%
k^{2}\ell ^{2}+2n}+\chi \frac{b_{0}\ell }{v_{F}}\right) .
\end{equation}%
We use the index $s=+1$($-1$) for the Landau levels that originate from the
conduction (valence) band of the Dirac cone and take $\hslash k$ as the
momentum of the electron in the direction of the magnetic field. At some
places in this paper, we use the alternative notation $n<0$ for the Landau
levels of the valence band. The corresponding eigenvectors are 
\begin{equation}
v_{_{\chi ,n,X,s}}\left( k,\mathbf{r}\right) =\frac{1}{\sqrt{L_{z}}}%
e^{ikz}\eta _{\chi ,n,X,s}\left( k,\mathbf{r}\right) ,
\end{equation}%
where $L_{z}$ is the length of the WSM\ in the $z$ direction and $\mathbf{r}$
a two-dimensional vector in the plane perpendicular to $z.$ The spinors are
given by%
\begin{equation}
\eta _{\chi ,0,X}\left( k,\mathbf{r}\right) =\left( 
\begin{array}{c}
0 \\ 
h_{0,X}\left( \mathbf{r}\right)%
\end{array}%
\right) ,
\end{equation}%
for the chiral level and by 
\begin{equation}
\eta _{\chi ,n,X,s}\left( k,\mathbf{r}\right) =\left( 
\begin{array}{c}
u_{\chi ,n,s}\left( k\right) h_{n-1,X}\left( \mathbf{r}\right) \\ 
v_{\chi ,n,s}\left( k\right) h_{n,X}\left( \mathbf{r}\right)%
\end{array}%
\right)  \label{landau}
\end{equation}%
for the others$.$ We set $h_{n,X}\left( \mathbf{r}\right) =0$ when $n<0.$
The $u^{\prime }s$ and $v^{\prime }s$ obey the normalization condition $%
\left\vert u_{\chi ,n,s}\left( k\right) \right\vert ^{2}+\left\vert v_{\chi
,n,s}\left( k\right) \right\vert ^{2}=1$ and are given by 
\begin{equation}
\left( 
\begin{array}{c}
u_{\chi ,n,s}\left( k\right) \\ 
v_{\chi ,n,s}\left( k\right)%
\end{array}%
\right) =\frac{1}{\sqrt{2}}\left( 
\begin{array}{c}
-\chi si\left( \sqrt{1+\chi s\frac{k\ell }{e_{n}\left( k\right) }}\right) \\ 
\sqrt{1-\chi s\frac{k\ell }{e_{n}\left( k\right) }}%
\end{array}%
\right) ,
\end{equation}%
where 
\begin{equation}
e_{n}\left( k\right) =\sqrt{k^{2}\ell ^{2}+2n}.
\end{equation}

In Eq. (\ref{landau}), the functions $h_{n,X}\left( \mathbf{r}\right) =\frac{%
1}{\sqrt{L_{y}}}\varphi _{n}\left( x-X\right) e^{-iXy/\ell ^{2}}$ are the
Landau level wave functions in the Landau gauge with $X$ the guiding-center
index and the $\varphi _{n}\left( x\right) ^{\prime }s$ are the wave
functions of the one-dimensional harmonic oscillator. Each Landau level $%
\left( \chi ,n,s,k\right) $ has degeneracy $N_{\varphi }=S/2\pi \ell ^{2}$
where $\ell =\sqrt{\hslash /eB}$ is the magnetic length and $S=L_{x}L_{y}$
is the area of the WSM\ perpendicular to the magnetic field.

The field operator can now be written as 
\begin{equation}
\Psi _{\chi }\left( \mathbf{r},z\right) =\frac{1}{\sqrt{L_{z}}}%
\sum_{n,s,X,k}e^{ikz}\eta _{\chi ,n,X,s}\left( k,\mathbf{r}\right) c_{\chi
,n,X,s,k}
\end{equation}%
and the second quantized form of the current operator is 
\begin{equation}
\mathbf{J}_{\chi }=\sum_{n,s,n^{\prime },s^{\prime },k}\sum_{X}\mathbf{%
\Gamma }_{\chi ;ns;n^{\prime }s^{\prime }}\left( k\right) c_{\chi
,n,X,s,k}^{\dag }c_{\chi ,n^{\prime },X,s^{\prime },k},  \label{Bcurrent}
\end{equation}%
with the matrix elements defined by%
\begin{eqnarray}
\Gamma _{\chi ;ns;n^{\prime }s^{\prime }}^{\left( x\right) }\left( k\right)
&=&-e\chi v_{F}\left[ u_{\chi ,n,s}^{\ast }\left( k\right) v_{\chi
,n-1,s^{\prime }}\left( k\right) \delta _{n-1,n^{\prime }}\right.
\label{gamxx} \\
&&\left. +v_{\chi ,n,s}^{\ast }\left( k\right) u_{\chi ,n+1,s^{\prime
}}\left( k\right) \delta _{n,n^{\prime }-1}\right] .  \notag \\
\Gamma _{\chi ;ns;n^{\prime }s^{\prime }}^{\left( y\right) }\left( k\right)
&=&ie\chi v_{F}\left[ u_{\chi ,n,s}^{\ast }\left( k\right) v_{\chi
,n-1,s^{\prime }}\left( k\right) \delta _{n-1,n^{\prime }}\right.
\label{gamyy} \\
&&\left. -v_{\chi ,n,s}^{\ast }\left( k\right) u_{\chi ,n+1,s^{\prime
}}\left( k\right) \delta _{n,n^{\prime }-1}\right] ,  \notag \\
\Gamma _{\chi ;ns;n^{\prime }s^{\prime }}^{\left( z\right) }\left( k\right)
&=&-e\chi v_{F}\left[ u_{\chi ,n,s}^{\ast }\left( k\right) u_{\chi
,n,s^{\prime }}\left( k\right) \right.  \label{gamzzz} \\
&&\left. -v_{\chi ,n,s,}^{\ast }\left( k\right) v_{\chi ,n,s^{\prime
}}\left( k\right) \right] \delta _{n,n^{\prime }}.  \notag
\end{eqnarray}

The current-current response tensor defined in Eq. (\ref{deux}) becomes%
\begin{eqnarray}
\chi _{\chi ,\left( \alpha ,\beta \right) }\left( \omega \right) &=&\zeta
\sum_{\substack{ k,n,n^{\prime },  \\ s,s^{\prime }}}\Gamma _{\chi
;n,s;n^{\prime },s^{\prime }}^{\left( \alpha \right) }\left( k\right) \Gamma
_{\chi ;n^{\prime },s^{\prime };n,s}^{\left( \beta \right) }\left( k\right)
\label{chid} \\
&&\times \frac{\left\langle n_{\chi ,n^{\prime },s^{\prime }}\left( k\right)
\right\rangle -\left\langle n_{\chi ,n,s}\left( k\right) \right\rangle }{%
\omega +i\delta -\left( E_{\chi ,n^{\prime }s^{\prime }}\left( k\right)
-E_{\chi ,n,s}\left( k\right) \right) /\hslash },  \notag
\end{eqnarray}%
where $\zeta =-1/2\pi \ell ^{2}\hslash L_{z},$ with $L_{z}$ the width of the
WSM\ in the $z$ direction. The filling factor for each Landau level is given
by 
\begin{equation}
\left\langle n_{\chi ,n,s}\left( k\right) \right\rangle =\frac{1}{N_{\varphi
}}\sum_{X}\left\langle c_{\chi ,n,X,s,k}^{\dag }c_{\chi
,n,X,s,k}\right\rangle .
\end{equation}%
The conductivity tensor is obtained from Eq. (\ref{un}) and the relative
dielectric tensor from Eq. (\ref{trois}).

The response functions $\chi _{\chi ,\left( \alpha ,z\right) }$and $\chi
_{\chi ,\left( z,\alpha \right) }$ with $\alpha =x,y$ are zero because of
the Kronecker deltas in the matrix elements $\mathbf{\Gamma }_{\chi
;ns;n^{\prime }s^{\prime }}.$ Moreover, for the interband part of the
response functions (i.e., $s\neq s^{\prime }$), we have the symmetry
relations $\chi _{\chi ,\left( x,x\right) }=\chi _{\chi ,\left( y,y\right) }$
and $\chi _{\chi ,\left( x,y\right) }=-\chi _{\chi ,\left( y,x\right) }.$
The optical selections rules imposed by the matrix elements are such that
only the dipolar transitions $\left\vert n\right\vert =\left\vert
n\right\vert \pm 1$ (here $n=0,\pm 1,\pm 2,...$) are permitted in $\chi
_{\chi ,\left( \alpha ,\beta \right) }$ with $\alpha _{,}\beta =x,y$ while
only the transitions $-n\rightarrow n$ are permitted in $\chi _{\chi ,\left(
z,z\right) }.$ (A tilt of a Weyl cone allows for a much richer interband
spectrum\cite{Parent2020,Goerbig2016}). We limit our study to the
ultra-quantum regime where the Fermi level lies in the chiral level. In this
case, intraband transitions can only occur in level $n=0$ and they
contribute only to $\chi _{\chi ,\left( z,z\right) }.$ The calculation of
the dynamic conductivity tensor including inter- and intra-Landau-level
contributions is summarized in Appendix B.

\section{MAXWELL EQUATIONS WITH THE AXION TERMS}

To compute the optical properties of a WSM, we must take into account the
modification of the Maxwell equations by the topological axion term $\theta
\left( \mathbf{r},t\right) =2\mathbf{b}\cdot \mathbf{r}-2b_{0}t.$ The
modified equations are\cite{Burkov2012}:%
\begin{eqnarray}
\nabla \cdot \mathbf{D} &=&\rho _{f}+\kappa \sqrt{\frac{\varepsilon _{0}}{%
\mu _{0}}}\mathbf{b}\cdot \mathbf{B,} \\
\nabla \cdot \mathbf{B} &=&0, \\
\nabla \times \mathbf{E} &=&-\frac{\partial \mathbf{B}}{\partial t}, \\
\nabla \times \mathbf{H} &=&\frac{\partial \mathbf{D}}{\partial t}+\mathbf{J}%
_{f}-\kappa \sqrt{\frac{\varepsilon _{0}}{\mu _{0}}}\left( \mathbf{b}\times 
\mathbf{E}-gb_{0}\mathbf{B}\right) \mathbf{,}
\end{eqnarray}%
where $\kappa =2\alpha /\pi $ with $\alpha $ the fine-structure constant. In
these equations, $\rho _{f}$ is the free (or induced) charge density and $%
\mathbf{J}_{f}=\overleftrightarrow{\mathbf{\sigma }}\cdot \mathbf{E}$ is the
induced current density with $\overleftrightarrow{\mathbf{\sigma }}$ the
conductivity tensor for the two-node model calculated in appendices A and B.
In this paper, we assume that the relative permeability $\mu _{r}=\mu /\mu
_{0}$ is unity.

In the context of the chiral magnetic effect which occurs in the
nonequilibrium condition where the chiral chemical potentials of both nodes
are different, the last term in the Ampere-Maxwell equation has $g=1.$ In
this paper, however, we have a different situation since we assume that the
two nodes are at equilibrium and that the two Dirac points are separated in
energy by $2\hslash b_{0}.$ The extra current $gb_{0}\mathbf{B}$ is then
related to the natural optical activity of the WSM and it follows that $%
g=1/3 $ in the dynamic case and $g=0$ in the static case%
\onlinecite{MaPesin2015,Zhong2016}. The gyrotropic effect of $b_{0}$ is, as
we will show, very small in comparison with that of axion term $\mathbf{b}$
so that, in the THz frequency range where we study the Kerr rotation, taking 
$g=1$ or $g=1/3$ or even $g=0$ does not make any numerical difference to the
results presented in this paper. In the small frequency limit, however,
keeping $g$ finite as $\omega \rightarrow 0$ leads to some strange results
as we will point out at some places in this paper.

The wave equation in the presence of the axion terms can be written in the
matrix form $M_{i,j}E_{j}=0$ where the components of the matrix $M$ are
given by%
\begin{equation}
M_{ij}=-c^{2}\left( q^{2}\delta _{ij}-q_{i}q_{j}\right) +\omega ^{2}%
\widetilde{\varepsilon }_{ij},  \label{matricem}
\end{equation}%
where $\delta _{ij}$ is the Kronecker delta. We have defined the tensor $%
\widetilde{\varepsilon }_{ij}$ by%
\begin{equation}
\widetilde{\varepsilon }_{ij}=\varepsilon _{ij}+\frac{i}{\omega }c\kappa
\varepsilon _{ijk}\left( b_{k}-\frac{gb_{0}}{\omega }q_{k}\right) ,
\label{epsitilde}
\end{equation}%
with $\varepsilon _{ijk}$ the antisymmetric Levi-Civita tensor and the
components of the dielectric tensor $\varepsilon _{ij}$ are given by%
\begin{equation}
\varepsilon _{ij}=\varepsilon _{b}\delta _{ij}+\frac{i}{\varepsilon
_{0}\omega }\sigma _{ij},  \label{eb}
\end{equation}%
where the electrons in each node are described by the hamiltonian of Eq. (%
\ref{hamil}) or Eq. (\ref{h5}) which contains the axion term $b_{0}$ but not 
$\mathbf{b}.$ In Eq. (\ref{eb}), $\varepsilon _{b}$ is the effective
background dielectric constant. We take $\varepsilon _{b}=1,$ but it must be
kept in mind that $\varepsilon _{b}$ can be substantially larger in WSMs.
For example $\varepsilon _{b}=6.2$ in type-I WSM TaAs\cite{Kotov2016}.

The dispersion relations and polarizations of the electromagnetic modes are
obtained from the equations $\det \left[ M\right] =0$ and $M_{i,j}E_{j}=0$
respectively. We stress once again that in the continuum model that we use,
the axion term $b_{0}$ is present in the dielectric tensor $%
\overleftrightarrow{\mathbf{\varepsilon }}$ \textit{and} in the Maxwell
equations but the axion term $\mathbf{b}$ appears in the modified Maxwell
equations but not in $\overleftrightarrow{\mathbf{\varepsilon }}$. The
induced charge density, if any, is given by%
\begin{equation}
\rho _{f}=i\varepsilon _{0}\mathbf{q}\cdot \left( \widetilde{\mathbf{%
\varepsilon }}\mathbf{\cdot E}\right) .  \label{induced}
\end{equation}

\section{KERR AND ELLIPTICITY ANGLES}

In this section, we derive the formulas necessary to compute the Kerr and
ellipticity angles. The external magnetic field, when present, is set along
the $z$ direction and we restrict our analysis to the case where the axion
term $\mathbf{b}=\mathbf{b}_{\Vert }=b_{z}\widehat{\mathbf{z}}$. We study
the effect of $\mathbf{b}_{\bot }=b_{x}\widehat{\mathbf{x}}+b_{y}\widehat{%
\mathbf{y}}$ in Sec. IX.

We consider a linearly polarized electromagnetic wave with amplitude $E_{0},$
polarization $\mathbf{e}_{0}$ and wave vector $\mathbf{q}_{0}$ arriving at
normal incidence on the surface of a semi-infinite WSM either along the $z$
axis with $\mathbf{e}_{0}\bot \mathbf{B}$ (the Faraday configuration) or
along the $x$ axis with $\mathbf{e}_{0}\cdot \mathbf{B\neq }0$ (the Voigt
configuration)\cite{Balkanski}. If $\mathbf{B}=0,$ we use the name Faraday
configuration for $\mathbf{e}_{0}\bot \mathbf{b}$ and Voigt configuration
for $\mathbf{e}_{0}\cdot \mathbf{b\neq }0$

\subsection{Faraday configuration}

In the Faraday configuration, the vacuum (medium 1)-WSM (medium 2) interface
is at $z=0.$ The electromagnetic field in medium 1 (with dispersion $%
q_{0}=\omega /c$) is given by

\begin{eqnarray}
\mathbf{E}_{1}\left( z\mathbf{,}t\right) &=&E_{0}\mathbf{e}%
_{0}e^{iq_{0}z}e^{-i\omega t}+E_{0}r\mathbf{e}_{r}e^{-iq_{0}z}e^{-i\omega t},
\label{e1} \\
\mathbf{B}_{1}\left( z\mathbf{,}t\right) &=&\frac{q_{0}}{\omega }E_{0}\left( 
\widehat{\mathbf{z}}\times \mathbf{e}_{0}\right) e^{iq_{0}z}e^{-i\omega t}
\label{e2} \\
&&-\frac{q_{0}}{\omega }E_{0}r\left( \widehat{\mathbf{z}}\times \mathbf{e}%
_{r}\right) e^{-iq_{0}z}e^{-i\omega t},  \notag
\end{eqnarray}%
while in medium 2, the two transmitted waves with wave vector $q_{1}$ and $%
q_{2}$ are given by 
\begin{eqnarray}
\mathbf{E}_{2}\left( z\mathbf{,}t\right) &=&E_{0}t_{1}\mathbf{e}%
_{1}e^{iq_{1}z}e^{-i\omega t}+E_{0}t_{2}\mathbf{e}_{2}e^{iq_{2}z}e^{-i\omega
t},  \label{e3} \\
\mathbf{B}_{2}\left( z\mathbf{,}t\right) &=&\frac{q_{1}}{\omega }%
E_{0}t_{1}\left( \widehat{\mathbf{z}}\times \mathbf{e}_{1}\right)
e^{iq_{1}z}e^{-i\omega t}  \label{e4} \\
&&+\frac{q_{2}}{\omega }E_{0}t_{2}\left( \widehat{\mathbf{z}}\times \mathbf{e%
}_{2}\right) e^{iq_{2}z}e^{-i\omega t}.  \notag
\end{eqnarray}%
In these equations, $\mathbf{e}_{0},\mathbf{e}_{r},\mathbf{e}_{1},\mathbf{e}%
_{2}$ are the (complex) polarization vectors for the incoming, reflected and
transmitted waves respectively and $r,t_{1},t_{2}$ are the reflection and
transmission factors.

At the surface of the WSM, the electric and magnetic fields must satisfy the
boundary conditions 
\begin{eqnarray}
\left( \mathbf{D}_{1}-\mathbf{D}_{2}\right) \cdot \widehat{\mathbf{n}}
&=&\rho _{free},  \label{eigen1} \\
\mathbf{E}_{1,\Vert }-\mathbf{E}_{2,\Vert } &=&0,  \label{eigen2} \\
\left( \mathbf{B}_{1}-\mathbf{B}_{2}\right) \cdot \widehat{\mathbf{n}} &=&0,
\label{eigen3} \\
\mathbf{B}_{1,\Vert }-\mathbf{B}_{2,\Vert } &=&0,  \label{eigen4}
\end{eqnarray}%
where the unit vector $\widehat{\mathbf{n}}$ points from medium $2$ to
medium $1.$

Defining the complex polarization vectors by%
\begin{equation}
\mathbf{e}_{j}=\alpha _{j}\widehat{\mathbf{x}}+\beta _{j}\widehat{\mathbf{y}}%
,  \label{veco}
\end{equation}%
with $j=0,r,1,2$ (a polarization component along $z$, if any, does not
contribute to the boundary conditions), we get from Eqs. (\ref{eigen1})-(\ref%
{eigen4}) the system of equations 
\begin{equation}
\left( 
\begin{array}{c}
r\alpha _{r} \\ 
r\beta _{r} \\ 
t_{1} \\ 
t_{2}%
\end{array}%
\right) =\left( 
\begin{array}{cccc}
1 & 0 & -\alpha _{1} & -\alpha _{2} \\ 
0 & 1 & -\beta _{1} & -\beta _{2} \\ 
0 & 1 & \frac{q_{1}}{q_{i}}\beta _{1} & \frac{q_{2}}{q_{i}}\beta _{2} \\ 
-1 & 0 & -\frac{q_{1}}{q_{i}}\alpha _{1} & -\frac{q_{2}}{q_{i}}\alpha _{2}%
\end{array}%
\right) ^{-1}\left( 
\begin{array}{c}
-\alpha _{0} \\ 
-\beta _{0} \\ 
\beta _{0} \\ 
-\alpha _{0}%
\end{array}%
\right) .  \label{coeff}
\end{equation}

In the Faraday configuration, the WSM has rotational symmetry along the $z$
axis since both $\mathbf{b}$ and $\mathbf{B}$ are along that axis$,$ and we
can take without loss of generality $\mathbf{e}_{0}=\widehat{\mathbf{x}}%
\allowbreak $ for the incident linear polarization. This choice implies $%
\alpha _{i}=1,\beta _{i}=0$ and the reflection and transmission factors are
then given by%
\begin{equation}
r\alpha _{r}=\frac{\left( 1-\frac{q_{1}}{q_{0}}\right) \left( 1+\frac{q_{2}}{%
q_{0}}\right) \alpha _{1}\beta _{2}-\left( 1+\frac{q_{1}}{q_{0}}\right)
\left( 1-\frac{q_{2}}{q_{0}}\right) \alpha _{2}\beta _{1}}{\left( \frac{q_{1}%
}{q_{0}}+1\right) \left( \frac{q_{2}}{q_{0}}+1\right) \left( \alpha
_{1}\beta _{2}-\alpha _{2}\beta _{1}\right) }  \label{r1}
\end{equation}%
and%
\begin{equation}
r\beta _{r}=-2\frac{\left( \frac{q_{1}}{q_{0}}-\frac{q_{2}}{q_{0}}\right)
\beta _{1}\beta _{2}}{\left( \frac{q_{1}}{q_{0}}+1\right) \left( \frac{q_{2}%
}{q_{0}}+1\right) \left( \alpha _{1}\beta _{2}-\alpha _{2}\beta _{1}\right) }%
.  \label{r2}
\end{equation}

With $b_{0},b_{z}\neq 0$, the matrix $M$ takes the block-diagonal form

\begin{equation}
M=\left( 
\begin{array}{ccc}
\omega ^{2}\widetilde{\varepsilon }_{xx}-c^{2}q^{2} & \omega ^{2}\widetilde{%
\varepsilon }_{xy} & 0 \\ 
-\omega ^{2}\widetilde{\varepsilon }_{xy} & \omega ^{2}\widetilde{%
\varepsilon }_{xx}-c^{2}q^{2} & 0 \\ 
0 & 0 & \omega ^{2}\widetilde{\varepsilon }_{zz}%
\end{array}%
\right) .
\end{equation}%
The dispersion relations of the electromagnetic waves are then given by (the
same dispersions are obtained in Refs. \onlinecite{Sa2021,Deng2021})%
\begin{eqnarray}
q_{1,\pm } &=&-\frac{g\kappa b_{0}}{2c}\pm \sqrt{\left( \frac{g\kappa b_{0}}{%
2c}\right) ^{2}+\frac{\omega ^{2}\xi _{-}}{c^{2}}},  \label{q11} \\
q_{2,\pm } &=&\frac{g\kappa b_{0}}{2c}\pm \sqrt{\left( \frac{g\kappa b_{0}}{%
2c}\right) ^{2}+\frac{\omega ^{2}\xi _{+}}{c^{2}}},  \label{q13}
\end{eqnarray}%
where we have defined the functions%
\begin{equation}
\xi _{\pm }=\varepsilon _{xx}\pm i\varepsilon _{xy}\mp \frac{c\kappa b_{z}}{%
\omega }.  \label{xsi}
\end{equation}%
The corresponding polarizations are given by%
\begin{equation}
\mathbf{e}_{1,\pm }=\frac{1}{\sqrt{2}}\left( 
\begin{array}{c}
i \\ 
1 \\ 
0%
\end{array}%
\right) ;\,\mathbf{e}_{2,\pm }=\frac{1}{\sqrt{2}}\left( 
\begin{array}{c}
-i \\ 
1 \\ 
0%
\end{array}%
\right) ,  \label{vec1}
\end{equation}%
while for the plasmon mode defined by $\varepsilon _{zz}=0,$ we have 
\begin{equation}
\mathbf{e}_{3}=\left( 
\begin{array}{c}
0 \\ 
0 \\ 
1%
\end{array}%
\right) .
\end{equation}%
The polarization vectors show that both the magnetic field (if present)\ and
the axion terms $b_{0},b_{z}$ lead to circular birefringence.

In deriving Eq. (\ref{coeff}), we have implicitly assumed that there are
only two electromagnetic modes propagating in each direction. This is not
obvious from the dispersions given by Eqs. (\ref{r1}) and (\ref{r2}) when $%
b_{0}\neq 0.$ We did check numerically, however, that such was the case for
the results presented in this paper.\textbf{\ }

From $\mathbf{e}_{1,+}$ and $\mathbf{e}_{2,+},$ we get $\alpha _{1}=-\alpha
_{2}=i/\sqrt{2}$ and $\beta _{1}=\beta _{2}=1/\sqrt{2}$ for the parameters
in Eqs. (\ref{r1}) and (\ref{r2}). To compute the Kerr angle, we first
define a function $\eta $ by%
\begin{equation}
\eta =\frac{r\beta _{r}}{r\alpha _{r}}=\frac{i\left( \frac{q_{1,+}}{q_{0}}-%
\frac{q_{2,+}}{q_{0}}\right) }{1-\frac{q_{1,+}q_{2,+}}{q_{0}^{2}}}.
\label{eta1}
\end{equation}%
In all our numerical results, we have checked by plotting the dispersions
that $q_{1,+}$ and $q_{2,+}$ always have positive real and imaginary parts
in the frequency range considered.

If $g=0$ in the $M,$ matrix (but $b_{0}$ is present in the conductivity
tensor), Eq. (\ref{eta1}) reduces to the simpler form%
\begin{equation}
\eta =\frac{r\beta _{r}}{r\alpha _{r}}=\frac{i\left( \sqrt{\xi _{-}}-\sqrt{%
\xi _{+}}\right) }{1-\sqrt{\xi _{-}\xi _{+}}}.  \label{eta2}
\end{equation}

In the general case where the polarization of the reflected wave is
elliptical, the Kerr angle $\theta _{K}$ is defined as the angle that the
major axis of the polarization ellipse makes with the direction of the
incident (linear) polarization i.e., the $x$ axis if $\mathbf{e}_{0}=%
\widehat{\mathbf{x}}.$ We use the definition\cite{BornWolf}%
\begin{equation}
\tan 2\theta _{K}=\frac{2\func{Re}\left[ \eta \right] }{1-\left\vert \eta
\right\vert ^{2}},  \label{kerrangle}
\end{equation}%
where $\theta _{K}\in \left[ -\pi /2,\pi /2\right] .$

The major and minor axis of the ellipse have length $a$ and $d$ respectively$%
.$ The ellipticity angle $\psi _{K}$ is defined\cite{BornWolf} as%
\begin{equation}
\tan \psi _{K}=\pm \frac{d}{a},
\end{equation}%
where the $\pm $ signs indicate the direction of rotation of the electric
field vector along the ellipse and $d/a\in \left[ 0,1\right] $. Thus, a
change in the sign of the ellipticity corresponds to a change in the
direction of the rotation of the electric field vector on the ellipse. The
ellipticity angle is obtained from the equation%
\begin{equation}
\sin 2\psi _{K}=\frac{2\func{Im}\left[ \eta \right] }{1+\left\vert \eta
\right\vert ^{2}}.  \label{ellipseangle}
\end{equation}%
A linear polarization has $\psi _{K}=0$ while a purely circular polarization
has $\psi _{K}=\pi /4.$ In the case of a WSM, $\left\vert \eta \right\vert $
is generally not small so that approximate formulas with $\left\vert \eta
\right\vert =0$ in the denominator of Eqs. (\ref{kerrangle}) and (\ref%
{ellipseangle}) cannot be used. We remark that, when $\func{Re}\left[ \eta %
\right] =0,$ $\theta _{K}=0$ if $1-\left\vert \eta \right\vert ^{2}>0$ but $%
\theta _{K}=\pi /2$ if $1-\left\vert \eta \right\vert ^{2}<0.$ An abrupt $%
\pi /2$ rotation of the Kerr angle is thus possible when $1-\left\vert \eta
\right\vert ^{2}$ changes sign with a small variation of $\omega $ or $B.$

\subsection{Voigt configuration}

In the Voigt configuration, the incident wave propagates along the $x$ axis
with the magnetic field along the $z$ axis. The analysis proceeds as in the
Faraday case but with $z$ replaced by $x$ in Eqs. (\ref{e1})-(\ref{e4}) and
with the polarization vectors of Eq. (\ref{veco}) now defined by%
\begin{equation}
\mathbf{e}_{j}=\alpha _{j}\widehat{\mathbf{z}}+\beta _{j}\widehat{\mathbf{y}}%
.
\end{equation}%
(A polarization component along $x$, if present, does not contribute to the
boundary conditions.) The reflection and transmission factors are still
given by Eq. (\ref{coeff}). In this configuration, it is necessary to take
the incident polarization at an angle with respect to the $y$ or $z$ axis in
order to get a Kerr rotation\cite{Balkanski}. We choose for the incident
polarization $\mathbf{e}_{0}=1/\sqrt{2}\left( \widehat{\mathbf{y}}+\widehat{%
\mathbf{z}}\right) $ and so we get 
\begin{eqnarray}
r\alpha _{r} &=&\frac{1}{\sqrt{2}}\frac{\left( 1-\frac{q_{1}q_{2}}{q_{0}^{2}}%
\right) }{\left( \frac{q_{1}}{q_{0}}+1\right) \left( \frac{q_{2}}{q_{0}}%
+1\right) } \\
&&-\frac{1}{\sqrt{2}}\frac{\left( \frac{q_{1}}{q_{0}}-\frac{q_{2}}{q_{0}}%
\right) \left( \alpha _{1}\beta _{2}+\alpha _{2}\beta _{1}-2\alpha
_{1}\alpha _{2}\right) }{\left( \frac{q_{1}}{q_{0}}+1\right) \left( \frac{%
q_{2}}{q_{0}}+1\right) \left( \alpha _{1}\beta _{2}-\alpha _{2}\beta
_{1}\right) }  \notag
\end{eqnarray}%
and%
\begin{eqnarray}
r\beta _{r} &=&\frac{1}{\sqrt{2}}\frac{\left( 1-\frac{q_{1}q_{2}}{q_{0}^{2}}%
\right) }{\left( \frac{q_{1}}{q_{0}}+1\right) \left( \frac{q_{2}}{q_{0}}%
+1\right) } \\
&&+\frac{1}{\sqrt{2}}\frac{\left( \frac{q_{1}}{q_{0}}-\frac{q_{2}}{q_{0}}%
\right) \left( \alpha _{1}\beta _{2}+\alpha _{2}\beta _{1}-2\beta _{1}\beta
_{2}\right) }{\left( \frac{q_{1}}{q_{0}}+1\right) \left( \frac{q_{2}}{q_{0}}%
+1\right) \left( \alpha _{1}\beta _{2}-\alpha _{2}\beta _{1}\right) }. 
\notag
\end{eqnarray}

In this configuration, the $M$ matrix with $b_{0},b_{z}\neq 0$ is given by

\begin{equation}
M=\left( 
\begin{array}{ccc}
\omega ^{2}\varepsilon _{xx} & \omega ^{2}\overline{\varepsilon }_{,y} & 0
\\ 
-\omega ^{2}\overline{\varepsilon }_{xy} & \omega ^{2}\varepsilon
_{xx}-c^{2}q^{2} & -ic\kappa gb_{0}q \\ 
0 & ic\kappa gb_{0}q & \omega ^{2}\varepsilon _{zz}-c^{2}q^{2}%
\end{array}%
\right) ,
\end{equation}%
where the function 
\begin{equation}
\overline{\varepsilon }_{xy}=\varepsilon _{xy}+i\frac{c\kappa b_{z}}{\omega }%
.  \label{epsibar}
\end{equation}

The dispersion relations of the electromagnetic waves are given by%
\begin{equation}
q_{i}=\pm \frac{\omega }{c}\sqrt{\frac{\varepsilon _{W}\pm \sqrt{\varepsilon
_{W}^{2}-4\varepsilon _{V}\varepsilon _{zz}}}{2}},  \label{combi}
\end{equation}%
where we have defined the dielectric functions 
\begin{eqnarray}
\varepsilon _{V} &=&\varepsilon _{xx}+\frac{\overline{\varepsilon }_{xy}^{2}%
}{\varepsilon _{xx}},  \label{epsiv} \\
\varepsilon _{W} &=&\varepsilon _{V}+\varepsilon _{zz}+\frac{\kappa
^{2}g^{2}b_{0}^{2}\allowbreak }{\omega ^{2}}.  \label{epsiw}
\end{eqnarray}%
The function $\varepsilon _{V}$ is the so-called Voigt dielectric function%
\cite{Balkanski}. The four corresponding eigenvectors are then (when $\omega
^{2}\varepsilon _{zz}-c^{2}q_{i}^{2}\neq 0,\varepsilon _{xx}\neq 0,b_{0}\neq
0$) 
\begin{equation}
\mathbf{e}_{i}=\frac{1}{\Lambda _{i}\left( q\right) }\left( 
\begin{array}{c}
-\frac{\overline{\varepsilon }_{xy}}{\varepsilon _{xx}} \\ 
1 \\ 
\frac{-i\kappa gb_{0}cq_{i}}{\omega ^{2}\varepsilon _{zz}-c^{2}q_{i}^{2}}%
\end{array}%
\right) ,  \label{f1}
\end{equation}%
with the normalization condition%
\begin{equation}
\Lambda \left( q_{i}\right) =\sqrt{1+\left\vert \frac{\overline{\varepsilon }%
_{xy}}{\varepsilon _{xx}}\right\vert ^{2}+\left\vert \frac{\kappa
gb_{0}cq_{i}}{\omega ^{2}\varepsilon _{zz}-c^{2}q_{i}^{2}}\right\vert ^{2}}.
\label{f2}
\end{equation}%
The wave vector $q_{i}$ in Eqs. (\ref{f1})-\ref{f2}) is one of the four
combinations given by Eq. (\ref{combi}). A particularity of the Voigt
configuration is that there is a component of the polarization vector along
the direction of propagation of the wave. It is given by%
\begin{equation}
E_{x}=-\frac{\overline{\varepsilon }_{xy}}{\varepsilon _{xx}}E_{y}.
\end{equation}

From the polarization vectors, we have for the function $\eta $ the
expression 
\begin{equation}
\eta =\frac{\left( \alpha _{1}-\alpha _{2}\right) \left( 1-\frac{q_{1}q_{2}}{%
q_{0}^{2}}\right) +\left( \alpha _{1}+\alpha _{2}-2\right) \left( \frac{q_{1}%
}{q_{0}}-\frac{q_{2}}{q_{0}}\right) }{\left( \alpha _{1}-\alpha _{2}\right)
\left( 1-\frac{q_{1}q_{2}}{q_{0}^{2}}\right) -\left( \alpha _{1}+\alpha
_{2}-2\alpha _{1}\alpha _{2}\right) \left( \frac{q_{1}}{q_{0}}-\frac{q_{2}}{%
q_{0}}\right) },  \label{etavoigt}
\end{equation}%
where 
\begin{equation}
\alpha _{i}=\frac{-i\kappa gb_{0}cq_{i}}{\omega ^{2}\varepsilon
_{zz}-c^{2}q_{i}^{2}},
\end{equation}%
where $i=1,2$ and $q_{1,}q_{2}$ are the two wave vectors of Eq. (\ref{combi}%
) propagating in the positive $x$ direction.

The $M$ matrix becomes block diagonal and the analysis simplifies
considerably if $g=0$ in the matrix $M.$ The four dispersions are then given
by 
\begin{eqnarray}
q_{1,\pm } &=&\pm \frac{\omega }{c}\sqrt{\varepsilon _{V}},  \label{q1v} \\
q_{2,\pm } &=&\pm \frac{\omega }{c}\sqrt{\varepsilon _{zz}},  \label{q2v}
\end{eqnarray}%
with the eigenvectors%
\begin{equation}
\mathbf{e}_{1,\pm }=\frac{1}{\sqrt{\left\vert \varepsilon _{xx}\right\vert
^{2}+\left\vert \overline{\varepsilon }_{xy}\right\vert ^{2}}}\left( 
\begin{array}{c}
-\overline{\varepsilon }_{xy} \\ 
\varepsilon _{xx} \\ 
0%
\end{array}%
\right) ,  \label{vo1}
\end{equation}%
and%
\begin{equation}
\mathbf{e}_{2,\pm }=\left( 
\begin{array}{c}
0 \\ 
0 \\ 
1%
\end{array}%
\right) .  \label{vo2}
\end{equation}%
The polarization vectors give $\beta _{1}=\varepsilon _{xx}/\sqrt{\left\vert
\varepsilon _{xx}\right\vert ^{2}+\left\vert \overline{\varepsilon }%
_{xy}\right\vert ^{2}},\beta _{2}=0$ and $\alpha _{1}=0,\alpha _{2}=1.$
Using these values in Eq. (\ref{coeff}) with $\mathbf{e}_{0}=\widehat{%
\mathbf{y}}$ or $\mathbf{e}_{0}=\widehat{\mathbf{z}}$ gives zero Kerr
rotation thus showing that it is necessary for the incident polarization to
make an angle with the $y$ or $z$ axis in order to get a non-zero result.
For the function $\eta ,$ we have

\begin{equation}
\eta =\frac{\left( 1-\sqrt{\varepsilon _{V}}\right) \left( 1+\sqrt{%
\varepsilon _{zz}}\right) }{\left( 1+\sqrt{\varepsilon _{V}}\right) \left( 1-%
\sqrt{\varepsilon _{zz}}\right) }.  \label{etav}
\end{equation}

The Kerr and ellipticity angles are given by Eq. (\ref{kerrangle}) and (\ref%
{ellipseangle}). In Eq. (\ref{kerrangle}), however, the angle is measured
with respect to the $z$ axis altough the real rotation angle is that
measured with respect to the incident polarization vector i.e., $\theta
_{K}-\pi /4$.

\section{KERR ROTATION IN A NON TOPOLOGICAL METAL}

For comparison with our results for a WSM, we give in this section the Kerr
angle for an ordinary (non topological) metal. In this case, the
conductivity tensor that enters the $M$ matrix is given by%
\begin{equation}
\sigma =i\varepsilon _{0}\omega _{p}^{2}\tau \left( 
\begin{array}{ccc}
\frac{\omega \tau +i}{\left( \omega \tau +i\right) ^{2}-\left( \omega
_{c}\tau \right) ^{2}} & \frac{-i\omega _{c}\tau }{\left( \omega \tau
+i\right) ^{2}-\left( \omega _{c}\tau \right) ^{2}} & 0 \\ 
\frac{i\omega _{c}\tau }{\left( \omega \tau +i\right) ^{2}-\left( \omega
_{c}\tau \right) ^{2}} & \frac{\omega \tau +i}{\left( \omega \tau +i\right)
^{2}-\left( \omega _{c}\tau \right) ^{2}} & 0 \\ 
0 & 0 & \frac{1}{\omega \tau +i}%
\end{array}%
\right) ,
\end{equation}%
where the magnetic field $\mathbf{B}=B\widehat{\mathbf{z}},$ $\tau $ is the
transport scattering time, $\omega _{c}=eB/m_{e}$ the cyclotron frequency
(with $m_{e}$ the electron mass) and $\omega _{p}=\sqrt{n_{e}e^{2}/m_{e}%
\varepsilon _{0}}$ (with $n_{e}$ the electronic density) is the plasmon
frequency given by $\varepsilon _{zz}\left( \omega _{p}\right) =0.$ For an
ordinary metal, $b_{0}=0$ and $\mathbf{b}=0.$

\begin{figure}
\centering\includegraphics[width = \linewidth]{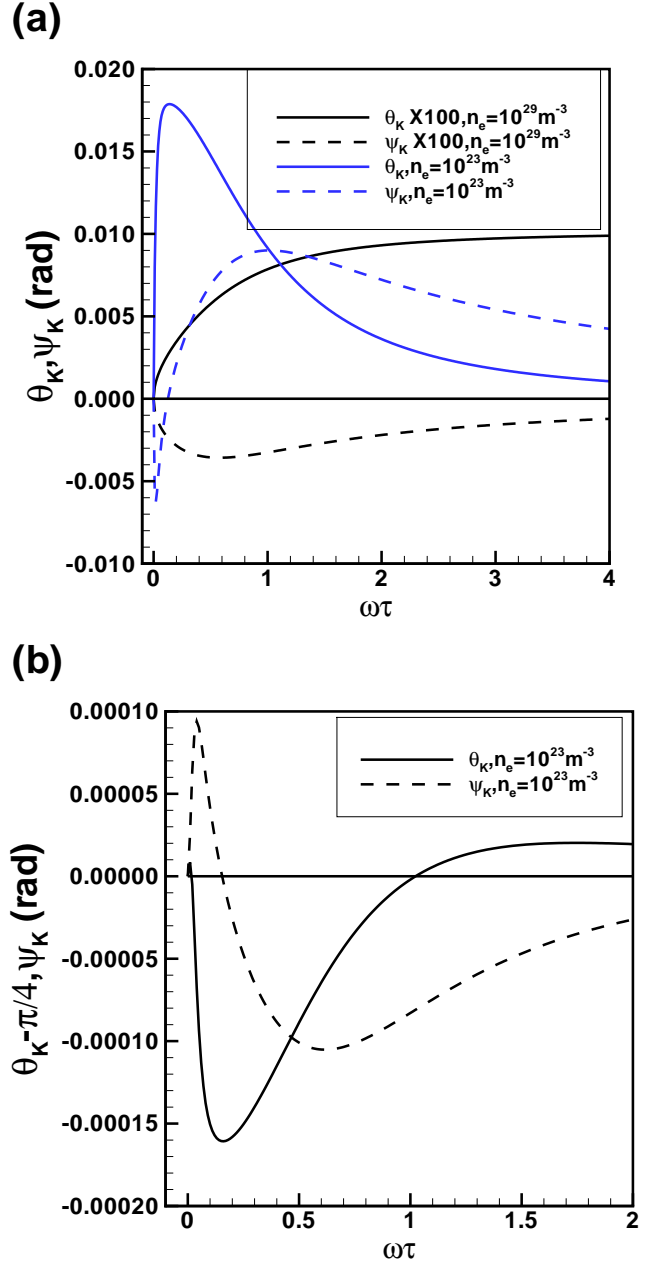} 
\caption{Frequency behavior of the Kerr and ellipticity angles $\protect\theta _{K},\protect\psi _{K}$ in the (a) Faraday and (b) Voigt
configuration for an ordinary (non topological) metal in a magnetic field.
Parameters: $\protect\tau =10^{-14}$ s, $B=10$ T, $\protect\omega _{c}\protect\tau =0.0178$ and $\protect\omega _{p}\protect\tau =178$ (if $n_{e}=10^{29}$ m$^{-3}$), $\protect\omega _{p}\protect\tau =0.178$ (if $n_{e}=10^{23}$ m$^{-3}$).}
\label{fig1}
\end{figure}

In the Faraday configuration, the dispersions, polarizations and function $%
\eta $ are given by Eqs. (\ref{q11}) and (\ref{q13}) and Eqs. (\ref{vec1})
and (\ref{eta2}). With a density $n_{e}\approx 10^{29}$ m$^{-3},$ and
relaxation time $\tau \approx 10^{-14}$ s, we get $\omega _{p}\tau \approx
1\,\allowbreak 78$ rad/s and $\omega _{c}\tau \approx 1.8\times 10^{-3}B$
with $B$ in Tesla. Figure 1 shows that, in the optical frequency range the
Kerr and ellipticity angles are extremely small. (The peak in $\theta _{K}$
occurs at the plasmon frequency in the ultraviolet frequency domain). If the
density is decreased to $n_{e}\approx 10^{23}$ m$^{-3}$ (a value closer to
what can be found is a WSM), the Kerr and ellipticity angles are
substantially increased. Equation (\ref{eta2}) shows that the Kerr angle
changes sign if the direction of the magnetic field is reversed.

In the Voigt configuration, the dispersions, polarizations and function $%
\eta $ are given by Eqs. (\ref{q1v}) and (\ref{vo2}) and Eq. (\ref{etav})
respectively. For $n_{e}=10^{29}$ m$^{-3},$ the Kerr $\,$($\theta _{K}-\pi
/4 $) and ellipticity angles are virtually zero ($<10^{-10}$ rad). For $%
n_{e}=10^{22}$ m$^{-3},$ both angles are substantially bigger but much
smaller than in the Faraday configuration for the same density. Since $%
\varepsilon _{V}$ is even in $B,$ the Kerr angle does not change sign if the
magnetic field is reversed. A particularity of that configuration is that in
the $q_{1,\pm }$ modes, $E_{x}=-\frac{\varepsilon _{xy}}{\varepsilon _{xx}}%
E_{y},$ so that there is a component of the polarization along the direction
of propagation. There is, however, from Eq. (\ref{induced}), no induced
charge.

The frequency profile $\theta _{K}\left( \omega \right) $ depends very much
on the relative value of $\omega _{p}$ and $\omega _{c}.$ Values of these
parameters other than those considered in this section are considered in
Ref. \onlinecite{Parent2020}.

\section{KERR ROTATION IN A WSM IN ZERO $\mathbf{B}$ FIELD}

We proceed to the study the frequency behavior of the Kerr and ellipticity
angles in a WSM in the absence of a magnetic field thus complementing
earlier works\cite{Kargarian,Sonowal2019} on this subject. We limit our
analysis to the case $b_{0},b_{z}\neq 0$. We discuss the effect of $\mathbf{b%
}_{\bot }$ in Sec. IX.

We take the wave vector $k$ and the axion term $b_{z}$ in units of $%
k_{a}=10^{8}~$m$^{-1}$ and choose $\hslash v_{F}k_{a}=19.7$ meV with $%
v_{F}=3\times 10^{5}$ m/s as our energy unit ($v_{F}k_{a}=3\times 10^{13}$ s$%
^{-1}$). In the presentation of our results, we use the dimensionless
variables $b_{0,r,}b_{z,r},k_{c,r},\omega _{r},k_{F,\chi ,r}$ which we
define by 
\begin{eqnarray}
b_{z} &=&b_{z,r}k_{a},k_{c}=k_{c,r}k_{a}, \\
b_{0} &=&b_{0,r}v_{F}k_{a},e_{F}=e_{F,r}\hslash v_{F}k_{a}, \\
k_{F,\chi } &=&k_{F,\chi ,r}k_{a},\,\omega =\omega _{r}v_{F}k_{a}.
\end{eqnarray}

We take these parameters in the range $\hslash b_{0},e_{F}\in \left[ 0,10%
\right] $ meV and $b_{z}\in \left[ -5\times 10^{8},5\times 10^{8}\right] $ m$%
^{-1},$ consistent with typical values reported in the literature\cite%
{Levy2020,Arnold2016} (for example, $b=3.2\times 10^{8}$ m$^{-1}$ in WTe$%
_{2} $\cite{Plie2017} and $v_{F}\approx 10^{6}$ m/s). In our dimensionless
units, these values correspond to $b_{0\,r},e_{F,r}\in \left[ 0,0.5\right] $
and $b_{z,r}\in \left[ -5,5\right] $. We take $\tau =10$ ps for the
intranode scattering time (the internode scattering time requires a large
momentum transfer and is substantially bigger\cite{Huang2021,Spivak}). We
take for the cutoff wave vector $k_{c,r}\approx 1000.$ The numerical values
of the Kerr angle, the position of its maximum, etc. all depend on the
precise choice of $k_{c},$ but not the qualitative aspects of frequency
profile $\theta _{K}\left( \omega \right) $.

\subsection{Faraday configuration}

In the Faraday configuration, the dispersions are obtained by setting $%
\varepsilon _{ij}=\varepsilon \delta _{ij}$ in Eq. (\ref{epsitilde}). The
dispersion relations, polarization and function $\eta $ are given by Eqs. (%
\ref{q11}) and (\ref{q13}) and Eqs. (\ref{vec1}) and (\ref{eta1})
respectively but with $\xi _{\pm }$ of Eq. (\ref{xsi}) replaced by $\xi
_{\pm }\left( \omega \right) =\varepsilon \left( \omega \right) \mp c\kappa
b_{z}/\omega .$

\begin{figure}
\centering\includegraphics[width = \linewidth]{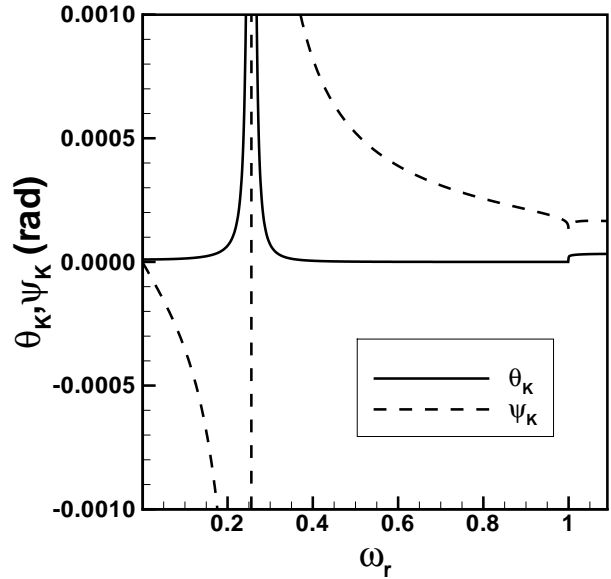} 
\caption{Effect of the parameter $b_{0,r} $ alone on the frequency behavior of the Kerr and ellipticity angles
in the Faraday configuration for a Weyl semimetal in zero magnetic field.
Parameters are $\protect\tau =10$ ps, $e_{F,r}=b_{z,r}=0,b_{0,r}=0.5,\mathbf{b}_{\bot }=0.$ The maximum of the Kerr angle (not shown) is $0.06$ rad.}
\label{fig2}
\end{figure}

The Kerr and ellipticity angles are shown in Fig. 2 for the case where we
artificially set $b_{z}=e_{F}=0$ in order to see the gyrotropic effect of $%
b_{0}$ alone. In this situation, there is an equal number of electrons and
holes but the WSM\ is at equilibrium. As can be seen from Fig. 2, the
gyrotropic effect of $b_{0}$ is very small with the exception of a narrow
peak at a frequency $\omega $ where $\varepsilon \left( \omega \right) =0$
i.e., at the plasmon frequency\cite{PlasmonBnul,Zhou2015}. The ellipticity
is also very small and changes sign at the plasmon frequency. The Kerr angle
is positive for $b_{0}>0,$ odd in $b_{0}$ and increases with $\left\vert
b_{0}\right\vert .$

Because of the Pauli blocking, the frequency threshold $\omega _{th,r}$ for
the electromagnetic absorption in each node is given by%
\begin{equation}
\omega _{th,r}=2\left\vert e_{F,r}-\chi b_{0,r}\right\vert .  \label{seuil}
\end{equation}%
Figure 2 shows that the Kerr angle has a step at this threshold which is at $%
\omega _{r}=2b_{0,r}=1$.

The dispersion relations in the Faraday configuration (Eqs. (\ref{q11}) and (%
\ref{q13}), contain both $b_{0}$ and $b_{z}$ in the ratio%
\begin{equation}
\frac{\frac{\omega ^{2}}{c^{2}}\frac{c\kappa b_{z}}{\omega }}{\left( \frac{%
\kappa b_{0}}{2c}\right) ^{2}}=\frac{4c\omega b_{z}}{\kappa b_{0}^{2}}=\frac{%
4\omega _{r}b_{z,r}}{\kappa b_{0,r}^{2}v_{F,r}}\approx 10^{6},
\end{equation}%
for $b_{z,r}=b_{0,r}=1,v_{F,r}=0.001$ and $\omega _{r}=1.$ Thus, in the THz
frequency range, the gyrotropic effect of $b_{0}$ is negligible in
comparison with that of $b_{z}$. This is also true in the Voigt
configuration although the dispersion relations are more complex to analyze
in this case.

Figure 3 shows the effect of $b_{z}$ on $\theta _{K}$ and $\psi _{K}$ when $%
b_{0}=e_{F}=0$. Although there is no carriers in this case, the Kerr angle
is not zero because of the presence of $b_{z}$ in $\xi _{\pm }$. With our
sign convention, $\theta _{K}$ is positive for negative $b_{z}$ and
vice-versa i.e., the Kerr effect is odd in $b_{z}$ as reported previously%
\cite{Sonowal2019}. Increasing $\left\vert b_{z}\right\vert $ does not
change the maximal value ($\theta _{K,\max }\approx 0.5$ rad at $\omega
_{\max ,r}$) but broadens the frequency range where the Kerr angle is
important and blueshifts the maximum. The Kerr angle is large in comparison
with that due to $b_{0}$ alone or to that in a normal metal (see Fig. 1).
When $b_{0}=0,$ the Kerr angle is even in $e_{F}$ so that electrons and
holes contribute the same way to the Kerr rotation. Consequently, the Kerr
angle is not zero at compensation when $e_{F}=0$ and $b_{0}\neq 0.$

In panel (a) of Fig. 3, the background dielectric constant is $\varepsilon
_{b}=1.$ Since $\varepsilon _{b}$ can be much larger in a WSM, we plot in
panel (b) the same curves but for $\varepsilon _{b}=10.$ We see that the
maximum value of the Kerr angle is unchanged as well as the overall profile
of the curves which are only compressed and pushed to lower frequencies. At
larger frequencies, the Kerr angle is reduced by a factor $\approx
\varepsilon _{b}.$

\begin{figure}
\centering\includegraphics[width = \linewidth]{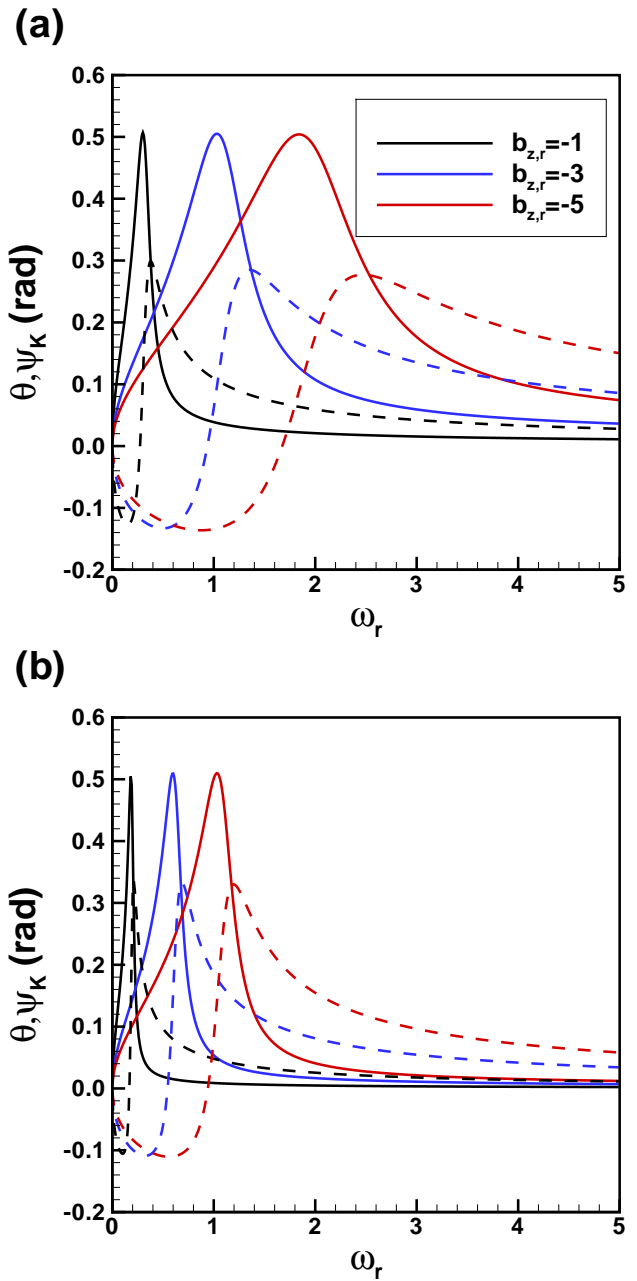} 
\caption{Frequency behavior of the Kerr
(full lines) and ellipticity angles (dashed lines) in the Faraday
configuration for a Weyl semimetal in zero magnetic field for several values
of the parameters $b_{z,r}$ and with $b_{0}=e_{F}=0.$ The background
dielectric function is $\protect\varepsilon _{b}=1$ in pannel (a) and $\protect\varepsilon _{b}=10$ in pannel (b).}
\label{fig3}
\end{figure}

Figure 4 shows the effect of $b_{z}$ on the Kerr angle for finite $b_{0}$
and $e_{F}.$ The threshold frequencies (which are independent of $b_{z}$)
for the absorption given by Eq. (\ref{seuil}) lead to two spikes in $\theta
_{K}\left( \omega \right) .$ Figure 4 shows how $\theta _{K}\left( \omega
\right) $ changes when the two threshold frequencies are on the left, in the
middle, and on the right of the frequency $\omega _{\max ,r}$ where the Kerr
angle is maximal. An interesting situation occurs when the two frequencies
are on the right of the maximum in $\theta _{K}.$ In this case, the Kerr
angle takes its maximal value $\theta _{K,\max }=\pi /2$ and a plateau where 
$\theta _{K}=0$ appears between the frequency $\omega _{r,-}$ where $\func{Re%
}\left[ \xi _{-}\left( \omega _{r,-}\right) \right] =0$ and the smaller of
the two frequencies $\omega _{th,r}.$ The width of this plateau increases
with decreasing $\left\vert b_{z}\right\vert $ since $\omega _{r,-}$ is then
pushed to lower frequencies. When this plateau is absent, $\omega _{r,-}$
coincides with the peak in the Kerr angle.

The Kerr angle is zero when $\eta $ is real (and $1-\left\vert \eta
\right\vert ^{2}>0$) or, equivalently when the wave vectors $q_{1+},q_{2+}$
are real so that the two electromagnetic waves propagate without
attenuation. As we argued above, the effect of $b_{0}$ is negligible in the
frequency range of Fig. 4 so that we can use Eq. (\ref{eta2}) instead of Eq.
(\ref{eta1}) to compute $\eta .$ The plateau where $\theta _{K}=0$ thus
occurs when the two dielectric functions $\xi _{\pm }\left( \omega \right) $
are real and positive as can be seen in Fig. 4.

The negative peak at small frequencies in Fig. 4 only occurs when the
intraband transitions are included in $\varepsilon \left( \omega \right) $
and when $b_{0}$ and/or $e_{F}$ are non-zero\textit{\ }so that there is a
finite density of electrons and/or holes. We find numerically, for the range
of parameters considered, that the position of this peak is at the frequency 
$\omega _{r,+}\left( \omega _{r,-}\right) $ where $\func{Re}\left[ \xi _{+}%
\right] =0\left( \func{Re}\left[ \xi _{-}\right] =0\right) $ when $%
b_{z}<0\left( b_{z}>0\right) .$ Its frequency is redshifted as $\left\vert
b_{z}\right\vert $ increases. In the opposite case, when $b_{z,r}\rightarrow
0,$ we have $\omega _{r,-}-\omega _{r,+}\rightarrow 0$ so that the negative
peak merges with the positive peak and the Kerr angle $\theta _{K}=0.$

\begin{figure}
\centering\includegraphics[width = \linewidth]{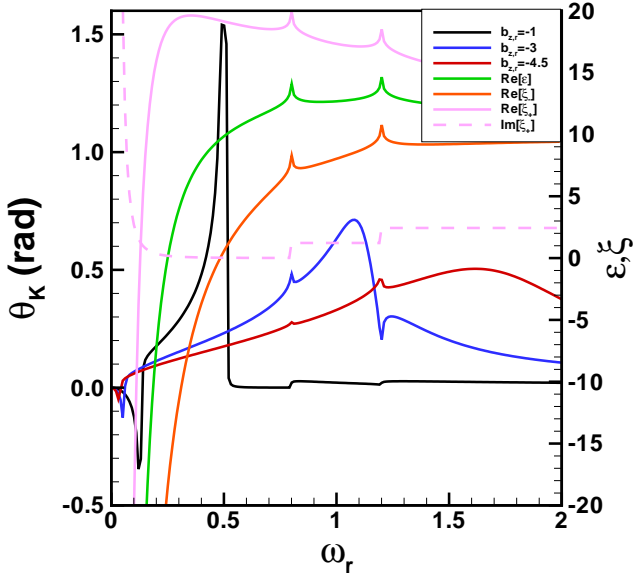} 
\caption{Kerr angle as a function of
frequency in the Faraday configuration for a Weyl semimetal in zero magnetic
field and for different values of $b_{z,r}$ and constant $b_{0,r}=0.1$ and $e_{F,r}=0.5.$ Also shown are the dielectric functions $\protect\varepsilon ,\protect\xi _{+},\protect\xi _{-}$ for $b_{z,r}=-1$. }
\label{fig4}
\end{figure}

\subsection{Voigt configuration}

In the Voigt configuration, the propagation vector and magnetic field are
respectively along the $x$ and $z$ axis. Setting $B=0$ and keeping only $%
b_{0}$ and $b_{z}$ in the $M$ matrix, the dispersion relations are given by
Eq. (\ref{combi}), the polarizations by Eq. (\ref{f1}) and $\eta $ by Eq. (%
\ref{etavoigt}) with $\varepsilon _{xx}=\varepsilon _{yy}=\varepsilon
_{zz}=\varepsilon .$ Since the light is incident on a surface with Fermi
arcs in this configuration (the Fermi arcs are present on surfaces parallels
to the Weyl node separation), their contribution to the conductivity should
in principle be taken into account. This contribution was estimated\cite%
{Kargarian} to be $\sigma _{xy}^{s}\simeq e^{2}\ln \left( 2b_{z}\lambda
\right) /\pi h$, where $\lambda $ is the wavelength of light. It is small in
comparison with the bulk contribution $\widetilde{\sigma }_{xy}=-i\widetilde{%
\varepsilon }_{xy}\varepsilon _{0}\omega =e^{2}b_{z}/\pi h$ in the THz range
where $\widetilde{\varepsilon }_{ij}$ is defined in Eq. (\ref{epsitilde}).
We thus neglect this contribution in our analysis.

Figure 5 shows the Kerr and ellipticity angles for a finite $b_{z,r}$ at $%
e_{F}=b_{0}=0.$ Contrary to the case of an ordinary metal, the Kerr rotation
angle $\theta _{K}-\pi /4$ is not negligible in the Voigt configuration. The
change of sign of $\theta _{K}-\pi /4$ occurs numerically at the frequency
where $\func{Re}\left[ \varepsilon _{V}\left( \omega \right) \right] =0$
which is also the frequency at which the ellipticity is maximal. The Kerr
angle is positive before the change of sign and is even\cite{Sonowal2019} in 
$b_{z}$. Its maximal value does not change much with $b_{z}$ but the range
of frequencies where the rotation is important increases with $\left\vert
b_{z}\right\vert .$ As in the Faraday case, there is a Kerr effect even in
the absence of carriers and even in the Pauli-blocked regime.

Figure 6 shows a more general result where $b_{z,r},b_{0,r}$ and $e_{F,r}$
are all finite and consequently the Kerr angle has more structure. The two
small spikes at frequency above $\omega _{r}=0.8$ correspond to the two
thresholds for the interband transitions given in Eq. (\ref{seuil}). Both
transitions are captured in the Kerr angle. The large discontinuities in the
Kerr angle correspond to transitions from $\pi /2$ to $-\pi /2.$ These two
values give the same angular position for the major axis of the ellipse.
There is in consequence no discontinuity in the rotating motion of the
polarization ellipse at these frequencies. Between these two peaks, the Kerr
angle is almost constant. The separation between the two peaks increases
with $\left\vert b_{z,r}\right\vert .$ Note that, with the exception of the
very small spikes at the interband threshold, the other two discontinuities
occur at $\omega _{r,1}$ and $\omega _{r,2}$ which are defined by $\func{Re}%
\left[ \varepsilon \left( \omega _{1,r}\right) \right] =0$ and $\func{Re}%
\left[ \varepsilon _{V}\left( \omega _{2,r}\right) \right] =0.$ The small
negative peak at small frequencies is again due to the intraband transitions
and is not present in the absence of carriers. We remark that increasing the
relaxation time above $\tau =10$ ps does not change at all the frequency
behavior of the Kerr angle shown in Fig. 6. Reducing $\tau ,$ however,
smooths all discontinuities at low $\omega _{r}$ as shown by the green curve
in Fig. 6 for which we took $\tau =0.1$ ps.

The first peak in the Kerr angle in Fig. 6 occurs at the plasmon frequency
as indicated by the red curve. In the limit $\omega \tau >>1$ and for $%
\omega <2v_{F}k_{F,\chi },$ Eq. (\ref{epsiwsm}) gives $\func{Im}\left[
\varepsilon \left( \omega \right) \right] \approx 0$ and 
\begin{equation}
\func{Re}\left[ \varepsilon \left( \omega \right) \right] \approx \beta
\left( \omega \right) -\frac{\alpha }{3v_{F}/c}\frac{4}{\pi \hslash ^{2}}%
\frac{\hslash ^{2}b_{0}^{2}+e_{F}^{2}}{\omega ^{2}},
\end{equation}%
where 
\begin{equation}
\beta \left( \omega \right) =1+\frac{\alpha }{6v_{F}/c}\sum_{\chi }\frac{1}{%
\pi }\ln \left( \left\vert \frac{\omega ^{2}-4v_{F}^{2}k_{c}^{2}}{\omega
^{2}-4v_{F}^{2}k_{F,\chi }^{2}}\right\vert \right) .
\end{equation}%
The plasmon frequency is then given by $\func{Re}\left[ \varepsilon \left(
\omega \right) \right] =0$ i.e., 
\begin{equation}
\omega _{p}=\sqrt{\frac{4\alpha }{3\pi \hslash ^{2}v_{F}/c}\frac{\hslash
^{2}b_{0}^{2}+e_{F}^{2}}{\beta \left( \omega \right) }},  \label{pplasmon}
\end{equation}%
consistent with the result given in Ref. \onlinecite{Zhou2015}. The
intraband plasmon frequency is red-shifted by the interband contribution
given by $\beta \left( \omega \right) .$ Equation (\ref{pplasmon}) can be
solved iteratively but a good approximation to the numerical value of the
plasmon frequency given in Fig. 6 is obtained by simply taking $\beta \left(
\omega =0\right) $ in Eq. (\ref{pplasmon}).

\begin{figure}
\centering\includegraphics[width = \linewidth]{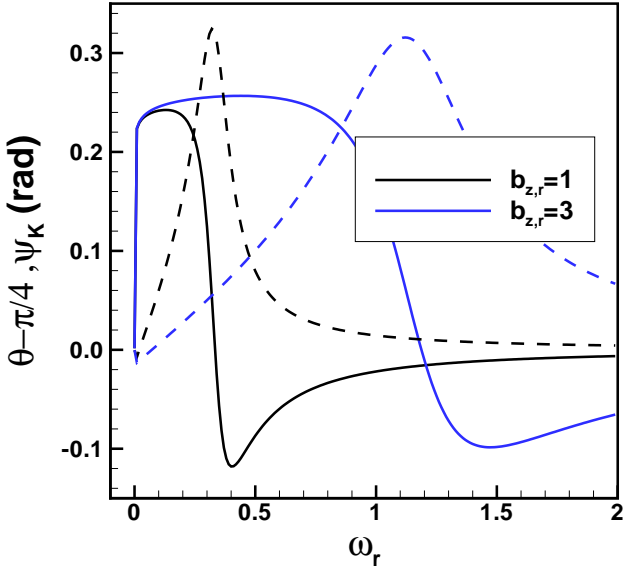} 
\caption{Frequency behavior of the Kerr
rotation angle $\protect\theta _{K}-\protect\pi /4$ (full lines) and
ellipticity angle (dashed lines) $\protect\psi _{K}$ in the Voigt
configuration for a Weyl semimetal in zero magnetic field for two values of $b_{z,r}$ and with $b_{0}=e_{F}=0.$}
\label{fig5}
\end{figure}

\begin{figure}
\centering\includegraphics[width = \linewidth]{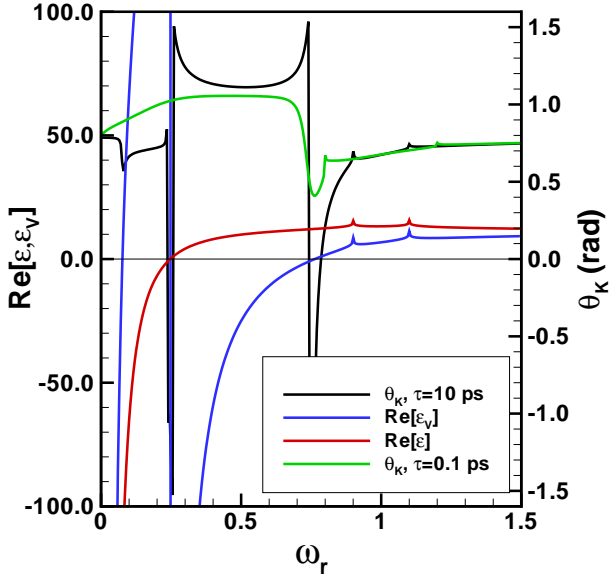} 
\caption{Frequency behavior of the Kerr
angle and real part of the dielectric functions $\protect\varepsilon _{V}\left( \protect\omega \right) $ and $\protect\varepsilon \left( \protect\omega \right) $ for a Weyl semimetal in zero magnetic field in the Voigt
configuration for the parameters $\protect\varepsilon _{F,r}=0.5,b_{z,r}=2,b_{0,r}=0.1$ and $\protect\tau =10$ ps$.$The green
curve gives the Kerr angle for $\protect\tau =0.1$ ps.} \label{fig6}
\end{figure}

\section{KERR EFFECT IN A WSM IN A FINITE MAGNETIC FIELD}

To study the effect of the magnetic field on the Kerr rotation, we define
the dimensionless variables 
\begin{eqnarray}
v_{F,r} &=&\frac{v_{F}}{c},b_{z,r}=b_{z}\ell ,  \notag \\
b_{0} &=&b_{0,r}v_{F}/\ell ,e_{F}=e_{F,r}\hslash v_{F}/\ell ,  \label{reduce}
\\
\omega &=&\omega _{r}v_{F}/\ell ,  \notag
\end{eqnarray}%
and choose $\hslash v_{F}/\ell =7.\,\allowbreak 69$ $v_{F,r}\sqrt{B}$ eV
with $B$ in Tesla as our unit of energy. The ratio $v_{F}/\ell =1.17\times
10^{16}v_{F,r}\sqrt{B}$ s$^{-1}$ so that the frequency $f=\omega /(2\pi
)=1.\,\allowbreak 86\times 10^{12}\omega _{r}\sqrt{B}$ Hz for $%
v_{F,r}=0.001. $ In terms of these dimensionless parameters, the range of
values considered in Sec. VII corresponds to $b_{0,r},e_{F,r}\in \left[
0,1.\,\allowbreak 3/\sqrt{B}\right] ,b_{z,r}\in \left[ -12.\,\allowbreak 83/%
\sqrt{B},12.\,\allowbreak 83/\sqrt{B}\right] $ for $v_{F,r}=0.001.$

We restrict our analysis to the ultraquantum limit where the Fermi level $%
e_{F}$ lies in the chiral Landau level of both nodes$.$ The Fermi wave
vector in each node, $k_{F,\chi },$ is then given by 
\begin{equation}
e_{F,r}=\chi \left( b_{0,r}-k_{F,\chi ,r}\ell \right)
\end{equation}%
and, in order for the $n\geq 1$ levels to be empty, the condition $%
e_{F,r}+b_{0,r}<\sqrt{2}$ must be satisfied. The occupation of the Landau
levels at $T=0$ K is given by Eqs. (\ref{occup1})-(\ref{occup3}) of Appendix
B. Since the density of states in one node is $g\left( E\right) =1/4\pi
^{2}\ell ^{2}\hslash v_{F},$ the Fermi level is given by%
\begin{equation}
e_{F,r}=2\pi ^{2}\ell ^{3}n_{e},
\end{equation}%
where $n_{e}$ is the total density of charge carriers in the WSM.

To describe the inter-Landau-level transitions, it is more convenient to use
negative values of $n$ for the Landau levels in the valence band ($s=-1$).
The threshold energy for the absorptive transitions in the $n=-1,0,1$ sector
are given by 
\begin{eqnarray}
\Delta _{0,1}^{\left( \chi \right) } &=&\sqrt{\left( e_{F,r}-\chi
b_{0,r}\right) ^{2}+2}-\left( e_{F,r}-\chi b_{0,r}\right) ,  \label{transi1}
\\
\Delta _{-1,0}^{\left( \chi \right) } &=&\sqrt{\left( e_{F,r}-\chi
b_{0,r}\right) ^{2}+2}+\left( e_{F,r}-\chi b_{0,r}\right) ,  \label{transi2}
\end{eqnarray}%
while for the dipolar transitions (with $n>0$) 
\begin{equation}
\Delta _{-n,n+1}^{\left( \chi \right) }=\sqrt{2\left\vert n\right\vert }+%
\sqrt{2\left( \left\vert n\right\vert +1\right) },  \label{dipolar}
\end{equation}%
and for the transitions captured by $\sigma _{zz}\left( \omega \right) $%
\begin{equation}
\Delta _{-n,n}^{\left( \chi \right) }=2\sqrt{2\left\vert n\right\vert }.
\end{equation}%
The transitions $\Delta _{0,1}^{\left( \chi \right) }$ $\left( \Delta
_{-1,0}^{\left( \chi \right) }\right) $ appear as peaks in the real part of
the conductivity $\sigma _{+}\left( \sigma _{-}\right) $ where $\sigma _{\pm
}=\sigma _{xx}\pm i\sigma _{xy},$ while the dipolar transitions $\Delta
_{-n,n+1}^{\left( \chi \right) }$ appear as peaks in both $\sigma _{+}$ and $%
\sigma _{-}.$ The transitions $\Delta _{-n,n}^{\left( \chi \right) }$
appears as peaks in $\sigma _{zz}$ only. The lowest frequency dipolar
transition has the threshold frequency $\omega _{r}=\Delta _{-1,2}^{\left(
\chi \right) }=\allowbreak 3.\,\allowbreak 41$ i.e., $f=0.32$ THz for $B=10$
T$.$ As in Sec. VII, we choose $\tau =10$ ps for the scattering time and
limit the summation over Landau levels to $n_{\max }=10$ unless indicated
otherwise. The qualitative aspect of $\theta _{K}\left( \omega \right) $ and 
$\chi _{K}\left( \omega \right) $ does not change significantly if we
increase $n_{\max }$ which is, in any case, limited by the bandwith of the
Weyl cones.

\subsection{Faraday configuration}

In the Faraday configuration$,$ the dispersion, polarizations and $\eta $
are given by Eqs. (\ref{q11}) and (\ref{q13}) and Eqs. (\ref{vec1}) and (\ref%
{eta1}) with $\xi _{\pm }$ defined in Eq. (\ref{xsi}). When $b_{0}=e_{F}=0,$
it follows that $\sigma _{xy}=0$ (since $x_{F}=0$ in Eqs. (\ref{sig1}) and (%
\ref{sig2})) and so $\varepsilon _{\pm }=\varepsilon _{xx}.$ If, in
addition, $b_{z}=0,$ then $\xi _{+}=\xi _{-}$ and $\eta =0.$ As expected the
magnetic field alone does not produce a Kerr rotation in the absence of
carriers. There is however a Kerr rotation when $b_{z}\neq 0$ and $%
e_{F}=b_{0}=0$ even if there is no carrier in the WSM. There is also a
rotation when $e_{F}=0$ and $b_{0}\neq 0$ i.e.,\textit{\ }at compensation
when the density of electrons is equal to the density of holes. Both
carriers contribute the same way to the Kerr angle as pointed out in Sec.
VII.

Figure 7 shows a typical result for the Kerr and ellipticity angles as well
as for the corresponding dielectric functions $\xi _{\pm }.$ We choose $%
b_{0,r}=0.2,b_{z,r}=0.6,e_{F,r}=1.$ We remark that $\xi _{-}$ has peaks at
the transitions $\Delta _{-1,0}^{\left( \chi \right) }$ while $\xi _{+}$
captures the transitions $\Delta _{0,1}^{\left( \chi \right) }$ for $\chi
=\pm .$ The dipolar transitions $\Delta _{-n,n+1}^{\left( \chi \right) }$
are present in both functions. In Fig. 7, the Kerr angle increases until it
reaches a plateau at $\theta _{K}=\pi /2$ in a small range of frequencies.
The onset of the plateau occurs when $\func{Re}\left[ \xi _{-}\left( \omega
\right) \right] =0.$ After the plateau, the Kerr angle drops abruptly to
zero and remains zero until the transition at $\omega _{r}=\Delta
_{0,1}^{\left( -\right) }.$ Discontinuities in $\theta _{K}\left( \omega
\right) $ occur at all four transitions in the $n=-1,0,1$ sector as well as
at all dipolar transitions$.$The transitions $\Delta _{-n,n}^{\left( \chi
\right) }$ are not captured by the Kerr angle in the Faraday configuration
since $\varepsilon _{zz}$ does not enter its calculation. Other transitions
will appear in $\theta _{K}\left( \omega \right) $ if the Weyl cones are
tilted\textbf{\ }since a tilt modifies the matrix elements of the current
operator and so the selection rules\cite{Goerbig2016,Parent2020}.

It may seem that there is a discontinuous change in the Kerr angle from $\pi
/2$ to $0$ at $\omega _{r}=0.34$ in Fig. 7. This is however not the case
since both the rotation and the ellipticity of the Kerr polarization change
with frequency. We illustrate this in Fig. 8 where we plot the trajectory of
the electric polarization vector for one cycle of oscillation at different
frequency points and for the parameters used in Fig. 7. The trajectory is an
ellipse with its major axis at $\theta _{K}=\pi /2$ at point 3 and becomes a
quasi circle at points $4$ and $5.$ This circle is then elongated along $%
E_{x}$ at points $6$ and $7$ and the major axis is now at $\theta _{K}=0.$
There has been, however, no discontinuous change in the trajectory of the
electric polarization vector in between these points.

\begin{figure}
\centering\includegraphics[width = \linewidth]{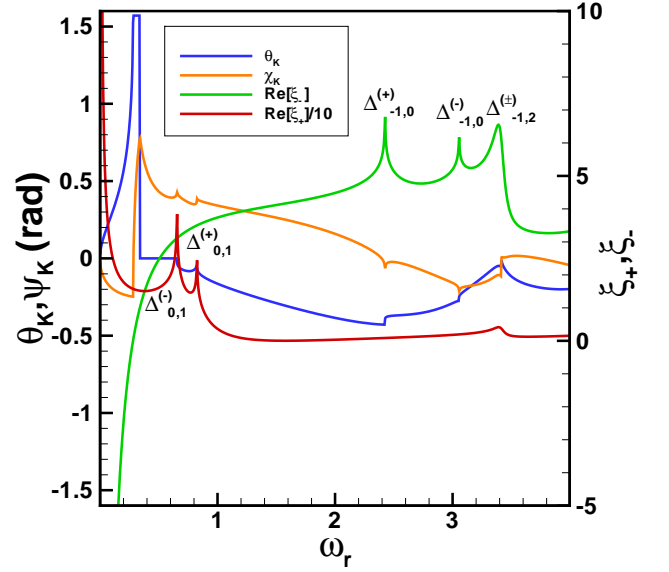} 
\caption{Frequency behavior of the Kerr
and ellipticity angles and real part of the dielectric functions $\protect\xi _{-},\protect\xi _{+-}$ for a Weyl semimetal in a magnetic field and in
the Faraday configuration. Parameters $e_{F,r}=1,b_{0,r}=0.2$ and $b_{z,r}=0.6.$ The $\Delta _{n,m}^{\left( \pm \right) }$ symbols indicate the
inter-Landau-level transitions.} \label{fig7}
\end{figure}

\begin{figure}
\centering\includegraphics[width = \linewidth]{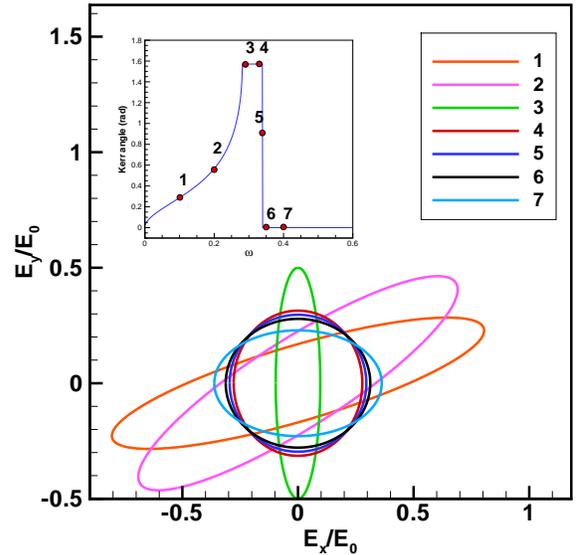} 
\caption{Trajectories of the
polarization vector for the different frequency points identified in the
inset. The sharp drop in the Kerr angle corresponds to a continuous change
in the shape of the ellipse. The parameters are those of Fig. 7.} \label%
{fig8}
\end{figure}

With $b_{z}=0$ and a finite Fermi level $e_{F,r}=1,$ the magnetic field
leads to a positive Kerr angle as shown by the black line in Fig. 9. With
our sign convention for $b_{z}$ and $b_{0},$ a positive value of $b_{z,r}$
leads to a negative Kerr angle in the absence of a magnetic field, as shown
in Sec. VII. Thus, at small frequencies increasing $b_{z,r}$ progressively
reduces the region where the Kerr angle induced by the magnetic field is
large. (In the opposite case where $b_{z,r}<0$, increasing $\left\vert
b_{z,r}\right\vert $ increases the region where the Kerr angle is large).
Further increasing $b_{z,r}$ changes the sign of the Kerr angle and
decreases it to zero. The plateaus where the Kerr angle is precisely zero or 
$\pi /2$ are in a frequency range where $\xi _{\pm }$ are both real and
positive so that $\eta $ is imaginary and $\tan \left( 2\theta _{K}\right)
=0 $ or $\pi $ depending on the sign of $1-\left\vert \eta \right\vert
^{2}<0 $ in Eq. (\ref{kerrangle})$.$ These two plateaus disappear when $%
b_{z,r}\leq 0$.

An interesting result of our calculation is that at precisely $%
b_{z,r}=e_{F,r},$ with $g=0$ in the matrix $M,$ the Kerr angle is zero in
the range $\omega _{r}\in \left[ 0,\Delta _{0,1}^{\left( -\right) }\right] $
i.e., up to the smallest of the four transition frequencies in the $n=-1,0,1$
sector. In this region, we find numerically that $\func{Re}[\eta ]=0$ and $%
1-\left\vert \eta \right\vert ^{2}>0$ in Eq. (\ref{kerrangle}) so that $%
\theta _{K}=0$. We do not observe this behavior when $B=0$. Indeed, this
result is due to the particular form of the conductivity tensor in the
extreme quantum limit i.e., to the fact that $\func{Re}\left[ \sigma _{xx}%
\right] =\func{Im}\left[ \sigma _{xy}\right] =0$ for $\omega _{r}<e_{F,r}.$
If we expand the conductivity $\sigma _{xx}$ and $\sigma _{xy}$ to first
order in $\omega _{r},$ we find that 
\begin{equation}
\xi _{\pm }=1-\frac{2\func{Im}\left[ \sigma _{xx}\right] }{\omega
\varepsilon _{0}}>0,  \label{condition}
\end{equation}%
so that $q_{1,+},q_{2,+}$ in Eqs. (\ref{q11}) and (\ref{q13}) are real and
positive and $\func{Re}[\eta ]=0$. The axion term $b_{z}$ in $\xi _{\pm }$
cancels the gyrotropic contribution from the $\varepsilon _{xy}$ in this
limit. Equation (\ref{condition}) is satisfied not just at small frequencies
but for $\omega _{r}\in \left[ 0,\Delta _{0,1}^{\left( -\right) }\right] ,$
something we were only able to check numerically. This observation can
provide a way to measure the axion term $b_{z}$ if the Fermi level is known.
When $g=1$ (or $g=1/3$), the Kerr angle is zero but only in the range $%
\omega _{r}\in \left[ \omega _{a,r},\Delta _{0,1}^{\left( -\right) }\right] $
where $\omega _{a,r}\approx 0.0001$ i.e., a very small frequency. Below $%
\omega _{a,r}$ we find that $1-\left\vert \eta \right\vert ^{2}<0$ and the
Kerr angle is $\theta _{K}=\pi /2.$ This remains true if $b_{z}=e_{F}=0$ so
that it would seem that the axion term $b_{0}$ causes a rotation of the
polarization from $\widehat{\mathbf{x}}$ to $\widehat{\mathbf{y}}$ at very
small frequency if $g$ is not zero, another curious (or spurious) effect of
a finite $b_{0}$ at $\omega \rightarrow 0.$

Another interesting fact when $\omega _{r}\in \left[ 0,\Delta _{0,1}^{\left(
-\right) }\right] $ and $b_{z,r}=e_{F,r},g=0$ is that the wave vectors $%
q_{1,\pm }$ and $q_{2,\pm }$ are real and positive and their dispersions are
linear in $q$ and gapless so that the two modes propagate without
attenuation as shown in Fig. 9(b). However, as soon as $b_{z,r}\neq e_{F,r},$
only one of the two modes remains gapless with a $\omega \sim q^{2}$
dispersion (if $b_{0}=0$)$.$ In the low-magnetic field limit where the
semi-classical Boltzmann equation is used to compute the conductivity
tensor, this mode would be the helicon mode discussed in Ref. %
\onlinecite{Helicon2015}.

When $b_{0}$ is finite Eqs. (\ref{q11}) and (\ref{q13}) give that the mode $%
q_{2,+},$ that propagates in the positive $z$ direction and $q_{1,-},$ a
mode propagating in the opposite direction have a finite wave vector $q\ell =%
\frac{\kappa b_{0}\ell }{c}=\kappa v_{F,r}\approx 10^{-5}$ at $\omega =0$.
This sort of instability in chiral matter and some of its consequences have
been discussed before\cite{Zebin2017}. In this reference however, $b_{0}$ is
associated with the chiral magnetic effect (CME) i.e., with a difference in
the chiral potential of the two nodes. In our case, the Fermi pockets are in
chemical equilibrium with a common Fermi level and $b_{0}$ is related to the
difference in energy between the two Dirac points, a property of the band
structure of the WSM. There is a gyrotropic current in this case\cite%
{MaPesin2015} but it should vanish in the static limit. In fact, according
to Ref. \onlinecite{Zhong2016}, the gyromagnetic contribution of $b_{0}$ and
its associate current in the Maxwell equations should be suppressed by
scattering at low frequency. If this is the case, this instability would
vanish. In any case, our results in the THz range are not affected by the
presence of this possible gap in the wave vector at very low frequency.

\begin{figure}
\centering\includegraphics[width = \linewidth]{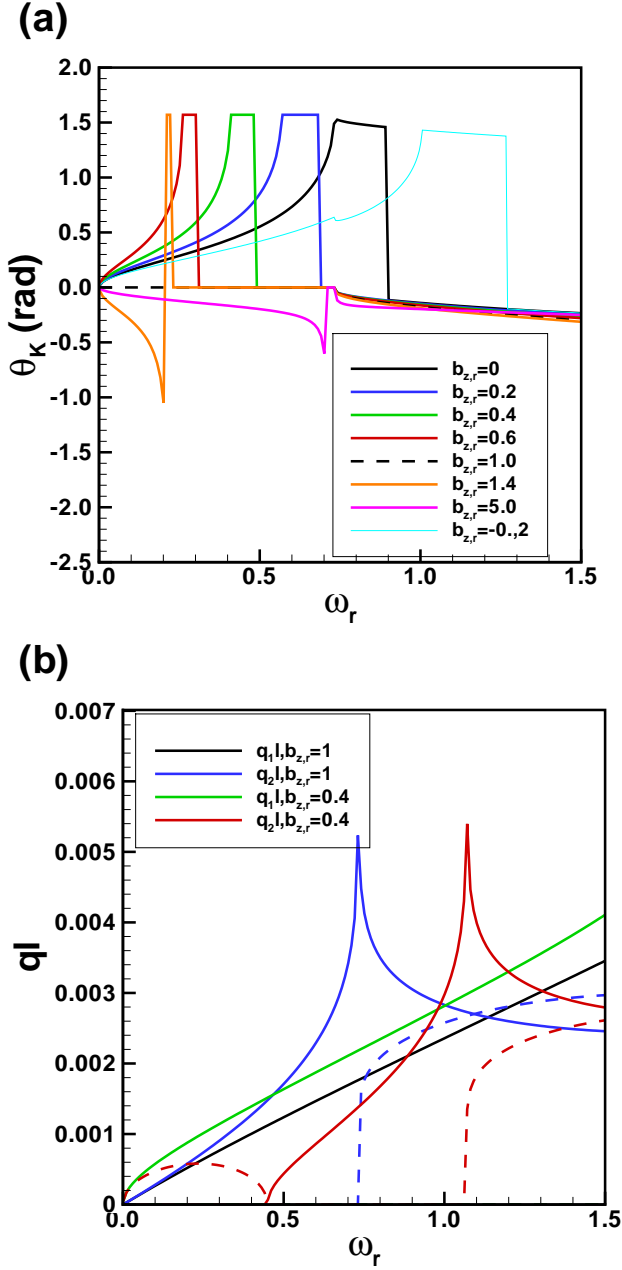} 
\caption{(a) Kerr angle as a function of
frequency for a Weyl semimetal in a magnetic field and in the Faraday
configuration for different values of $b_{z,r}$ and with $e_{F,r}=1,b_{0,r}=0 $ and $n_{mzx}=20.$ (b) Real (full lines) and imaginary
part (dashed lines) of the wave vector $q\ell $ for $e_{F,r}=1,b_{0,r}=0$
and $b_{z,r}=0.4,1.0$.} \label{fig9}
\end{figure}

\subsection{Voigt configuration}

We consider an incident wave propagating along the $x$ axis with the
external magnetic field again in the $z$ direction. The dispersion relations
are given by Eq. (\ref{combi}), the polarizations by Eq. (\ref{f1}) and $%
\eta $ by Eq. (\ref{etavoigt}). As for the $B=0$ case, the same numerical
results are obtained by using the simpler formulas with $g=0$ ($b_{0}$ must
be kept in the dielectric tensor, however).

Figure 10 shows $\theta _{K}\left( \omega \right) $ for the same parameters
as those used in Fig. 9 for the Faraday configuration. Since $\varepsilon
_{zz}$ now enters in the calculation of the Kerr angle and because the
intra-Landau level transitions in $n=0$ are included in $\varepsilon _{zz},$
we must also specify the magnetic field and the relaxation time in addition
to the dimensionless variables $b_{0,r},b_{z,r},e_{F,r}$ and $\omega _{r}.$
The Kerr angle has a plateau at $\theta _{K}=\pi /4$ (i.e., no effective
rotation of the polarization) at small frequencies before dropping abruptly
to $\theta _{K}=-\pi /4.$ As in the Faraday case, small spikes signal the
dipolar transitions $\Delta _{-1,2}^{\left( \chi \right) }=3.42,$ $\Delta
_{0,1}^{\left( \chi \right) }=0.73$ and $\Delta _{-1,0}^{\left( \chi \right)
}=2.73$. A unique feature of the Voigt configuration is that the transitions 
$\Delta _{-n,n}^{\left( \chi \right) }$ are now present in $\theta
_{K}\left( \omega \right) $. Increasing $b_{z,r}$ decreases the width of the
plateau where $\theta _{K}=\pi /4$ and so increases the range of frequencies
where the effective rotation angle is important. Contrary to what we found
in the Faraday configuration, $\theta _{K}$ is not zero when $%
b_{z,r}=e_{F,r} $ but its frequency profile is distinctively different , at
low frequency, than the profiles for $b_{z,r}\neq e_{F,r}.$

\begin{figure}
\centering\includegraphics[width = \linewidth]{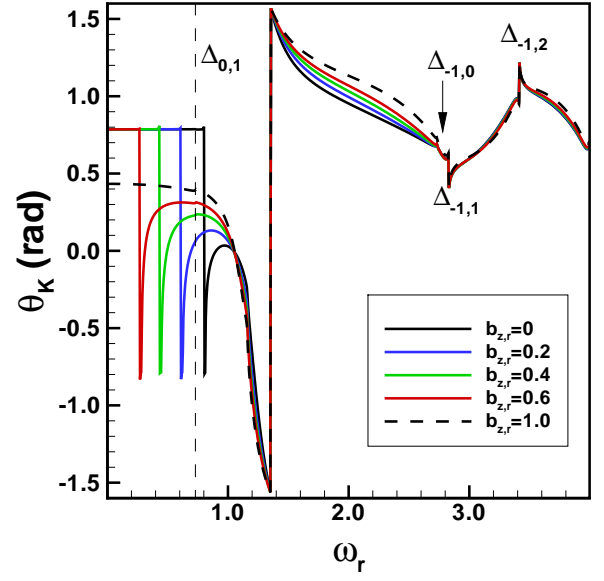} 
\caption{Kerr angle as a function of
frequency for a Weyl semimetal in a magnetic field and in the Voigt
configuration for different values of $b_{z,r}$ and with $e_{F,r}=1,b_{0,r}=0,B=10$ T, $\protect\tau =10$ ps. The rotation angle is $\protect\theta _{K}-\protect\pi /4.$ The Landau level transitions indicated
by the $\Delta _{n,m}$ symbols have the same value for both nodes since $b_{0}=0.$}
\label{fig10}
\end{figure}

Another feature of the Voigt configuration is that $\theta _{K}$ has a
change of curvature and a zero ellipticity at a frequency where $\varepsilon
_{zz}\left( \omega \right) =0$ i.e., at the plasmon frequency $\omega _{p}$
as indicated in Fig. 11. This feature is also present in Fig. 10 but
difficult to see because of the superposition of the different curves. If we
consider only the intraband transitions, we have from $\varepsilon
_{zz}\left( \omega \right) =0$ in the limit $\omega \tau >>1$ that the
plasmon frequency is given by

\begin{equation}
\omega _{p}=\sqrt{\frac{e^{3}v_{F}B}{2\pi ^{2}\hslash ^{2}\varepsilon _{0}},}
\end{equation}%
which is the known result\cite{Spivak} for two nodes. The plasmon frequency
is independent of the relaxation time in this limit. The $\omega _{p}\sim 
\sqrt{B}$ behavior is an experimental signature of a WSM\cite{Sa2021} and of
the chiral anomaly. This frequency is corrected by inter-Landau-level
transitions and scattering. The more accurate result is found by solving the
self-consistent equation 
\begin{equation}
\omega _{p}=\sqrt{\frac{2e^{3}v_{F}B}{4\pi ^{2}\varepsilon _{0}\hslash ^{2}}%
\frac{1}{1-\frac{8\alpha c\Lambda \left( \omega _{p,r}\right) }{\pi v_{F}}}-%
\frac{1}{\tau ^{2}}},  \label{plasma}
\end{equation}%
where%
\begin{equation}
\Lambda \left( \omega _{r}\right) =\sum_{n>0}n\int_{0}^{+\infty }dx\frac{1}{%
\left( x^{2}+2n\right) ^{3/2}}\frac{1}{\omega _{r}^{2}-4\left(
x^{2}+2n\right) }.
\end{equation}%
A very good approximation of the plasmon frequency found in Fig. 11 is
obtained by taking $\omega _{r}=0$ in $\Lambda \left( \omega _{r}\right) .$
The inter-Landau-level transitions decrease the plasmon frequency because $%
\Lambda <0$ when $\omega _{r}<\sqrt{2}$. Figure 11 also shows that the end
of the plateau where $\theta _{K}=\pi /4$ and the negative peak leading to $%
\theta _{K}=-\pi /4$ are at a frequency corresponding to $\func{Re}\left[
\varepsilon _{V}\left( \omega \right) \right] =0$ as in the $B=0.$ The
negative peak, however, does not disappear if $e_{F}=b_{0}=0$ and is not due
to the intraband transitions.

\begin{figure}
\centering\includegraphics[width = \linewidth]{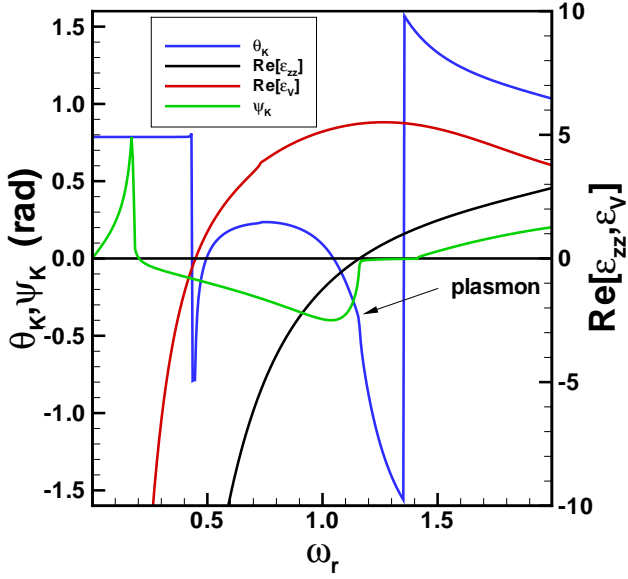} 
\caption{Frequency behavior of the Kerr
and ellipticity angles and of the real part of the dielectric functions $\protect\varepsilon _{zz}$ and $\protect\varepsilon _{V}$ for a Weyl
semimetal in a magnetic field and in the Voigt configuration. The actual
rotation of the polarization is $\protect\theta _{K}-\protect\pi /4.$
Parameters: $b_{z,r}=0.2,e_{F,r}=b_{0,r}=0,B=10$ T and $\protect\tau =10$
ps. The kink in the Kerr angle is at the plasmon frequency.} \label{fig11}
\end{figure}

\section{EFFECT OF $\mathbf{b}_{\bot }$ ON THE KERR ANGLE}

We study in this section the effect of a perpendicular component $\mathbf{b}%
_{\bot }$ on the Kerr angle. To obtain analytical results that are
manageable, we set $g=0$ in the matrix $M$ but keep $b_{0}$ in the
conductivity tensor. The magnetic field is again along the $z$ axis and the
light propagates along that axis. The $M$ matrix is

\begin{equation}
M=\left( 
\begin{array}{ccc}
\omega ^{2}\varepsilon _{xx}-c^{2}q_{z}^{2} & \omega ^{2}\varepsilon
_{xy}+i\omega c\kappa b_{z} & -i\omega c\kappa b_{y} \\ 
-\omega ^{2}\varepsilon _{xy}-i\omega c\kappa b_{z} & \omega ^{2}\varepsilon
_{xx}-c^{2}q_{z}^{2} & i\omega c\kappa b_{x} \\ 
i\omega c\kappa b_{y} & -i\omega c\kappa b_{x} & \omega ^{2}\varepsilon _{zz}%
\end{array}%
\right) .
\end{equation}%
With the axion term in spherical coordinates%
\begin{equation}
\mathbf{b}=b\sin \theta \cos \varphi \widehat{\mathbf{x}}+b\sin \theta \sin
\varphi \widehat{\mathbf{y}}+b\cos \theta \widehat{\mathbf{z}},
\end{equation}%
the dispersion relations and corresponding polarizations are given by%
\begin{equation}
q_{\pm }^{2}=-\frac{\kappa ^{2}}{2\varepsilon _{zz}}\Lambda _{\pm }\left(
\theta ,\varphi =0,b\right) ,
\end{equation}%
where%
\begin{eqnarray}
\Lambda _{\pm }\left( \theta ,\varphi ,b\right) &=&b^{2}\sin ^{2}\theta \cos
\left( 2\varphi \right) \\
&&\mp \sqrt{b^{4}\sin ^{4}\theta +\frac{4\omega ^{2}\varepsilon _{zz}^{2}}{%
c^{2}\kappa ^{2}}\left( b\cos \theta -i\frac{\omega \varepsilon _{xy}}{%
c\kappa }\right) ^{2}}  \notag
\end{eqnarray}%
and (for $b\neq 0$)%
\begin{equation}
\mathbf{e}_{\pm }=\frac{1}{\sqrt{\left\vert E_{x,\pm }\right\vert
^{2}+\left\vert E_{z,\pm }\right\vert ^{2}+1}}\left( 
\begin{array}{c}
E_{x,\pm } \\ 
1 \\ 
E_{z,\pm }%
\end{array}%
\right) ,
\end{equation}%
where we have defined 
\begin{equation}
E_{x,\pm }=\frac{2i\frac{\omega \varepsilon _{zz}}{c\kappa }\left( b\cos
\theta -i\frac{\omega \varepsilon _{xy}}{c\kappa }\right) +b^{2}\sin
^{2}\theta \sin 2\varphi }{\Lambda _{\pm }\left( \theta ,\varphi ,b\right) },
\label{Ex}
\end{equation}%
and (for $\varepsilon _{zz}\neq 0$) 
\begin{equation}
E_{z,\pm }=-ib\sin \theta \frac{E_{x,\pm }\sin \varphi -\cos \varphi }{\frac{%
\omega \varepsilon _{zz}}{c\kappa }}.  \label{ez}
\end{equation}%
There is a component of the polarization in the direction of propagation of
the wave when $\theta \neq 0.$

If the polarization of the incident wave is along the $x$ axis, the function 
$\eta $ that enters in the calculation of the Kerr angle is given by (with $%
E_{y}=1$)%
\begin{equation}
\eta =\frac{-2\left( \frac{q_{+}}{q_{0}}-\frac{q_{-}}{q_{0}}\right) }{\left(
1-\frac{q_{+}}{q_{0}}\right) \left( 1+\frac{q_{-}}{q_{0}}\right)
E_{x,+}-\left( 1+\frac{q_{+}}{q_{0}}\right) \left( 1-\frac{q_{-}}{q_{0}}%
\right) E_{x,-}},  \label{etab}
\end{equation}%
where $q_{0}=\omega /c.$ Since, the rotational symmetry in the $x-y$ plane
is broken, the Kerr angle changes with the azimuthal angle $\varphi $ and
depends on the orientation of the polarization vector of the incident wave.

Figure 12 shows the Kerr angle for several orientations of the axion term $%
\mathbf{b}$ for $e_{F,r}=1,\left\vert \mathbf{b}_{r}\right\vert =0.4$ and $%
b_{0,r}=0.$ The frequency domain where $\theta _{K}$ is large increases with
the polar angle $\theta $ if $e_{F}$ is not zero. This is consistent with
the behavior shown in Fig. 9 since a finite $\theta \in \left[ 0,\pi /2%
\right] $ decreases $b_{z,r}$ when $\left\vert \mathbf{b}_{r}\right\vert $
is fixed$.$ When $\varphi =0,\pi /2$ and $\theta \neq 0,$ a new peak appears
on the right of the plateau where $\theta _{K}=\pi /2$ and increases in
amplitude with $\theta \in \left[ 0^{+},\pi /2\right] .$ The frequency at
which $\theta _{K}=0$ on the left side of this peak is the plasmon
frequency. Indeed, when, $\func{Re}\left[ \varepsilon _{zz}\right] =0$, $%
\func{Im}\left[ \varepsilon _{zz}\right] \approx 0,\theta =\pi /2$ and $%
\varphi =0,\pi /2$, Eq. (\ref{Ex}) becomes $E_{x,\pm }=0$ so that $\eta
\rightarrow \infty $ and $\theta _{K}=0.$

We found previously that a nonzero $b_{z}$ gives a Kerr angle even in the
absence of carriers. In the present case, if $\theta =\pi /2,\varphi =0$ and 
$e_{F}=b_{0}=0$ (no carriers), the Kerr angle is zero at all frequencies
(not shown in Fig. 12).

\begin{figure}
\centering\includegraphics[width = \linewidth]{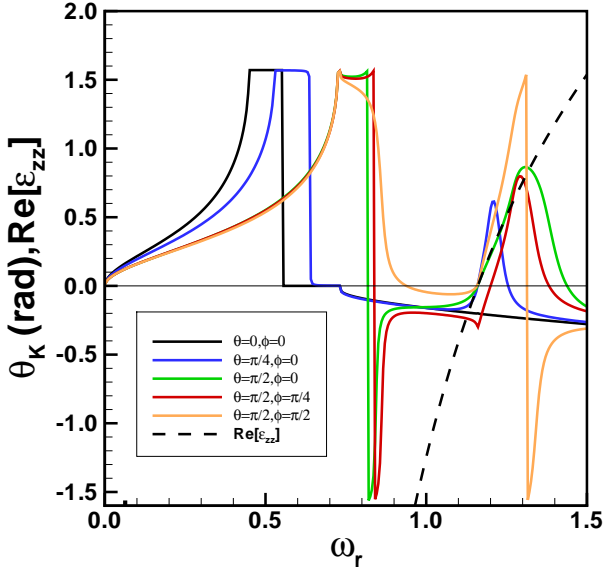} 
\caption{Kerr angle as a function of
frequency for a Weyl semimetal in a magnetic field and in the Faraday
configuration for different orientations of the vector $\mathbf{b}$ and with 
$e_{F,r}=1,b_{0,r}=0,b=0.4$ and real part of the dielectric function $\protect\varepsilon _{zz}$ near $\protect\varepsilon _{zz}=0.$}
\label{fig12}
\end{figure}

Figure 9 shows that when $b_{z,r}=e_{F,r}$ and $g=0,$ the Kerr angle is zero
in the frequency range $\omega _{r}\in \left[ 0,\Delta _{0,1}^{\left(
-\right) }\right] .$ This remains true if $\mathbf{b}$ is tilted with $b_{z}$
kept constant at $b_{z,r}=e_{F,r}$ and $\varphi =0$.

\section{MAGNETIC FIELD BEHAVIOR OF THE KERR ANGLE}

We end our study of the Kerr rotation in a WSM\ by considering the situation
where the frequency $f$ is kept fixed and the magnetic field is varied. We
restrict our analysis to the Faraday configuration. Figure 13 shows the Kerr
angle for $f=5$ THz, $b=5\times 10^{8}$ m$^{-1}$, $\varphi =0$, $%
v_{F}=3\times 10^{5}$ m/s, $n_{e}=10^{21}$ m$^{-3},\tau =10$ ps and $%
b_{0}=0. $ Note that the dimensionless parameters $b_{r},b_{0,r},\omega
_{r},e_{F,r}$ that enter the calculation of the conductivity tensor and the $%
M$ matrix change with magnetic field. In addition the Fermi level $e_{F}$
decreases with increasing $B$ when the density is fixed. To observe any
transition in $\theta _{K}\left( B\right) $, we must ensure that $f$ is
chosen such that $\omega _{r}=2\pi f/(v_{F}/\ell )$ goes through the
different inter-Landau-level transitions when the magnetic field is varied.
In addition, in order to capture the plasmon mode, $\omega _{r}$ must
intersect the frequency $\omega _{p,r}$ where $\varepsilon _{zz}\left(
\omega _{p,r}\right) =0$ and $\theta $ must be finite. Finally, the
condition $e_{F,r}+b_{0,r}<\sqrt{2}$ must be satisfied at all $B$ in order
for the WSM\ to stay in the ultra-quantum limit where our analysis is valid.
The magnetic field range $B\in \left[ 0.5,6\right] $ T with the above
parameters satisfy these conditions.

The black curve in Fig. 13 is the function $\theta _{K}\left( B\right) $ for
the special case where $b=0$ thus showing the effect of the magnetic field
alone. We see that $\theta _{K}\left( B\right) =0$ when the magnetic field
is such that $\omega _{r}<\Delta _{-1,0}$ which is the lowest-energy
transition when the Fermi level is above the Dirac point$.$ At lower $B$,
the transitions $\Delta _{-1,2}$ and $\Delta _{0,1}$ are captured by the
Kerr angle. The transition $\Delta _{-1,1}$ is captured only if $\theta \neq
0$ because it is only then that $\varepsilon _{zz}\left( \omega \right) $,
which contains these transitions, enters in the equation for the Kerr angle.

When $\theta $ is non-zero, a downward turn appears in $\theta _{K}\left(
B\right) $ above the magnetic field $B^{\ast }$ for which $\varepsilon
_{zz}\left( \omega _{p,r}\right) =0$ (note that, in our reduced units, $%
\omega _{p,r}$ is independent of $B$ in the clean limit). For $B>B^{\ast },$
Fig. 13 shows that $\left\vert \theta _{K}\left( B\right) \right\vert $
increases with $\theta $ and the shoulder is pushed to higher $B.$ The Kerr
angle is effectively zero from $B^{\ast }$ to the shoulder and tends to a
constant at large $B.$

\begin{figure}
\centering\includegraphics[width = \linewidth]{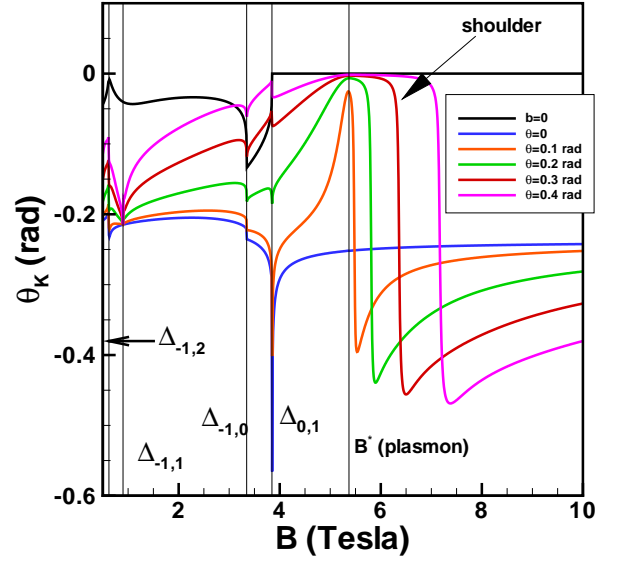} 
\caption{Behavior of the Kerr angle with
magnetic field at fixed frequency $f=5$ THz in the Faraday configuration for
different inclination angles $\protect\theta $ of the axion term $\mathbf{b}. $ The parameters are $b=5\times 10^{8}$ m$^{-1}$, $n_{e}=10^{21}$ m$^{-3},\protect\tau =10$ ps, $v_{F}=3\times 10^{5}$ m/s,$\protect\varphi =0$ and $b_{0}=0.$ The black curve is for $b=0$. }
\label{fig13}
\end{figure}

The Fermi level $e_{F,r}\sim 1/B^{3/2}$ and frequency $\omega _{r}\sim
1/B^{1/2}$ so that, in the limit of very large $B,$ $\varepsilon
_{xy}\rightarrow 0$ and 
\begin{eqnarray}
\varepsilon _{xx}^{\ast } &\equiv &\varepsilon _{xx}\left( B\rightarrow
\infty \right) \\
&=&1+\frac{\kappa }{2v_{F,r}}\left( 1+2\sum_{n>1}\int_{-\infty }^{+\infty }dx%
\frac{\Lambda _{n}\left( x\right) }{\Delta _{n}^{3}\left( x\right) }\right) ,
\notag
\end{eqnarray}%
where $\Lambda _{n}\left( x\right) $ and $\Delta _{n}\left( x\right) $ are
defined in Appendix B. The constant $\varepsilon _{xx}^{\ast }$ is real and
positive. We thus have, from Eqs. (\ref{xsi}) and (\ref{eta2},) that%
\begin{equation}
\xi _{\pm }=\varepsilon _{xx}^{\ast }\mp \frac{c\kappa b_{z}}{\omega }
\end{equation}%
and the function $\eta $ that enters the formula for the Kerr angle is, for $%
\theta =0$ and $b_{0}=0,$ given by 
\begin{equation}
\eta =\frac{i\left( \sqrt{\varepsilon _{xx}^{\ast }+\frac{c\kappa b_{z}}{%
\omega }}-\sqrt{\varepsilon _{xx}^{\ast }-\frac{c\kappa b_{z}}{\omega }}%
\right) }{1-\sqrt{\varepsilon _{xx}^{\ast 2}-\left( \frac{c\kappa b_{z}}{%
\omega }\right) ^{2}}}.
\end{equation}%
The Kerr angle in this limit is plotted in Fig. 14 as a function of $b_{z}$
for $f=5$ THz and $B=20$ T. On the positive axis, the Kerr angle is zero at
small $b_{z}$ where $1-\left\vert \eta \right\vert ^{2}>0$ and changes
abruptly to $\theta _{K}=-\pi /2$ when $1-\left\vert \eta \right\vert ^{2}<0$
[see Eq. (\ref{kerrangle})]. As $b_{z}$ increases further, $\left\vert
\theta _{K}\right\vert $ decreases. The end of the plateau where $\theta
=-\pi /2$ is at $b_{z}^{\ast }=\omega \varepsilon _{xx}^{\ast }/c\kappa .$
Note that, since $\varepsilon _{xy}\rightarrow 0,$ the magnetic field has no
effective rotating power at large $B.$ Figure 14 shows that the Kerr angle
is odd in $b_{z}$ as expected in the Faraday configuration.

\begin{figure}
\centering\includegraphics[width = \linewidth]{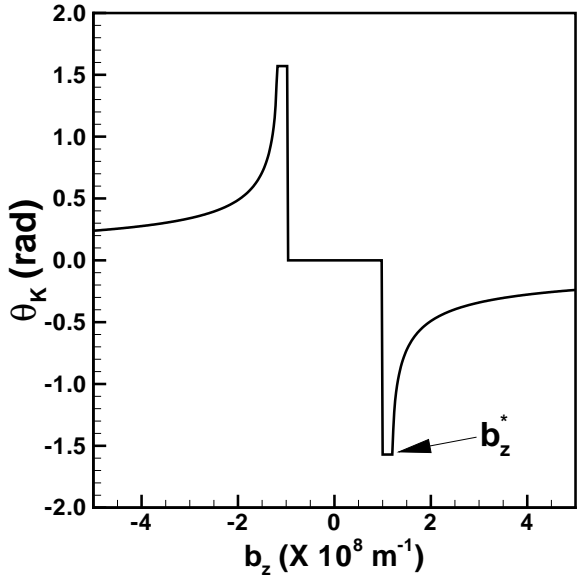} 
\caption{Kerr angle as a function of the
axion term $b_{z}$ in the limit of large magnetic field and for $B=20$ T, $f=5$ THz, $\protect\theta =\protect\varphi =0,$ $n_{e}=10^{21}$ m$^{-3},$ $v_{F}=3\times 10^{5}$ m/s, $b_{0}=0.$}
\label{fig14}
\end{figure}

\section{CONCLUSION}

We have studied the effect of the axion terms $b_{0}$ and $\mathbf{b}$ on
the magnetic field and frequency profile of the Kerr angle using a simple
two-node model of a Weyl semimetal with broken time reversal and/or
inversion symmetries. We have considered the effect of the axion terms alone
and then in the presence of an external static magnetic field. In both
cases, our study was made in the Faraday and Voigt configurations. When a
magnetic field was considered, we analyzed the Kerr rotation in the
ultra-quantum regime where the Fermi level lies in the chiral Landau level.
For the case $\mathbf{B}=0,$ our study complements earlier works on the
subject while the $\mathbf{B}\neq 0$ case, with the axion terms, has not
been studied before.

The Kerr angle has a rich frequency behavior which depends sensitively on
the value of the different parameters of the WSM. We find that it can be
quite large, reaching its maximum value $\theta _{K}=\pm \pi /2$ at certain
frequencies. The gyrotropic power of $\mathbf{b}$ is much bigger than that
of $b_{0}$ which, in our calculations, has mainly the effect of unbalancing
the electronic density in both nodes. Depending on its orientation with
respect to the external magnetic field, the axion term $\mathbf{b}$ can
increase or decrease the Kerr rotation caused by the magnetic field.
Moreover, the axion term $\mathbf{b},$ contrary to the magnetic field $%
\mathbf{B},$ enforces a Kerr rotation even when the Fermi level $e_{F}=0$
(i.e., no electronic density in both nodes if $b_{0}=0$ or equal density of
electrons and holes if $b_{0}\neq 0$).

Allowed electronic transitions are captured by the frequency or magnetic
field profile of the Kerr angle$.$ The Faraday and Voigt configurations
capture a different set of transitions. In the Voigt configuration and with $%
\mathbf{B}\neq 0,$ the frequency of the plasmon mode is signaled by a small
kink in $\theta _{K}\left( \omega \right) .$ The plasmon mode also appears
in $\theta _{K}\left( \mathbf{B}\right) $ if the vector $\mathbf{b}$ is
slightly tilted with respect to the external magnetic field.

In the ultra-quantum limit, the particular form of the dielectric function
of the Weyl semimetal allows the Kerr angle to be zero in a large range of
frequencies in the Faraday configuration. More interestingly, this range can
extend all the way down to $\omega =0$ in the particular case where the
component of the axion term along the magnetic field satisfies the condition 
$b\hslash v_{F}=e_{F}$ (for $\mathbf{b}\Vert \mathbf{B}$) and the axion term 
$b_{0}$ is absent from the Maxwell equations.

There are very few measurements of the Kerr angle in WSMs reported in the
literature. Most of them measure the Faraday rotation. In Ref. %
\onlinecite{Okamura2020}, however, a large Kerr rotation of $\theta
_{K}=0.056$ rad was measured in the magnetic WSM Co$_{3}$Sn$_{2}$S$_{2},$
which hosts three pairs of Weyl nodes\cite{Wang2017}, in a weak magnetic
field at a photon energy of $\hslash \omega =0.1$ eV (in the infrared
spectrum). This angle is large in comparison with that measured in
ferromagnetic metals.

In this paper, we used the simplest hamiltonian possible for a time-reversal
symmetry broken WSM in order to isolate the effect of the axion terms $b_{0}$
and/or $b_{z}$ on the Kerr rotation. To our knowledge, there is a present no
known WSM\ with this idealized electronic dispersion and a single pair of
Weyl nodes. Most WSMs have several pairs or nodes with a certain amount of
tilt, anisotropy, and nonlinearity in the dispersion. From a previous study
by one of us\cite{Parent2020} where the axion terms were not considered, we
know that the Kerr rotation, in a magnetic field, is also dependent on the
tilt of each node. It is, of course, possible to extend the approach used in
our paper to a WSM with an arbitrary number of pairs of nodes (possibly
tilted) by calculating the total conductivity tensor if the parameters of
each node as well as their position in energy with respect to the commun
Fermi level are known. It would be difficult, then to isolate the
contribution of the axion terms to the Kerr rotation. 

Magnetic Weyl semimetals with a single pair of Weyl nodes are hard to find
but recent $ab$ $initio$ calculations\cite{Wang2019,Soh2019} and ARPES\cite%
{Soh2019}\ measurements have shown that EuCd$_{2}$As$_{2}$ in a magnetic
field greater than $B_{c}=1.6$ T applied along its $c$ axis is a Weyl
semimetal with a single pair of tilted Weyl nodes. In this WSM, $b_{0}=0$
and $b_{z}=2.6\times 10^{8}$ m$^{-1}$ and $b_{z}$ is found to be almost
constant for $B>B_{c}.$ The magnetic Weyl semimetal K$_{2}$Mn$_{3}$(AsO$_{4}$%
)$_{3}$ is also predicted by first-principles calculations to host a single
pair of nodes\cite{Nie2022}. In both cases, however, the low-energy
hamiltonian is more complex than the one used in our paper. Moreover, Kerr
rotation measurements on these model Weyl semimetals have not yet been done.

\begin{acknowledgments}
R. C. was supported by a grant from the Natural Sciences and Engineering
Research Council of Canada (NSERC).
\end{acknowledgments}

\appendix

\section{CONDUCTIVITY TENSOR OF THE WSM FOR$\ \mathbf{B}=0$\ }

The conductivity of a Weyl node is the sum of the interband and intraband
contributions. The interband term is obtained by calculating the
current-current response function in the clean limit as given by Eq. (\ref%
{chic}). At $T=0$ K, using the results of Sec. II, we find%
\begin{eqnarray}
\sigma _{\chi ,inter}\left( \omega \right) &=&\frac{e^{2}\omega }{12hv_{F}}%
\left[ \theta \left( \omega -2v_{F}\left\vert k_{F,\chi }\right\vert \right)
\right. \\
&&\left. -\frac{i}{\pi }\ln \left( \left\vert \frac{\omega
^{2}-4v_{F}^{2}k_{c}^{2}}{\omega ^{2}-4v_{F}^{2}k_{F,\chi }^{2}}\right\vert
\right) \right] ,  \notag
\end{eqnarray}%
where $v_{F}k_{c}$ is a an ultraviolet cutoff and $k_{F,\chi }$ is the Fermi
wave vector in each node which is related to $b_{0}$ and the Fermi level $%
e_{F}$ by Eq. (\ref{quatre}).

For the intraband conductivity, it is necessary to include disorder and the
retarded current-current response is then given by%
\begin{eqnarray}
\chi _{\chi }^{\left( xx\right) }\left( \omega \right) &=&\frac{%
e^{2}v_{F}^{2}}{V\hslash }\sum_{s}\sum_{\mathbf{k}}\left\vert \eta _{\chi
,s}^{\dag }\left( \mathbf{k}\right) \sigma ^{\left( x\right) }\eta _{\chi
,s^{\prime }}\left( \mathbf{k}\right) \right\vert ^{2} \\
&&\times \int_{-\infty }^{+\infty }d\omega ^{\prime }\int_{-\infty
}^{+\infty }d\omega ^{\prime \prime }A_{\chi ,s}\left( \mathbf{k},\omega
^{\prime }\right) A_{\chi ,s}\left( \mathbf{k},\omega ^{\prime \prime
}\right)  \notag \\
&&\times \frac{F\left( \omega ^{\prime \prime }\right) -F\left( \omega
^{\prime }\right) }{\omega +i\delta +\omega ^{\prime \prime }-\omega
^{\prime }},  \notag
\end{eqnarray}%
where $F\left( \omega \right) =1/\left( e^{\beta \hslash \omega }+1\right) $
with $\beta =1/k_{B}T$ and $A_{\chi ,s}\left( \mathbf{k},\omega \right) $ is
the spectral weight of the disorder averaged Green's function $\left\langle
G_{\chi ,s}^{R}\left( \mathbf{k},\omega \right) \right\rangle $ which is
given by 
\begin{equation}
\left\langle G_{\chi ,s}^{R}\left( \mathbf{k},\omega \right) \right\rangle
=\int_{-\infty }^{+\infty }d\omega \frac{A_{\chi ,s}\left( \mathbf{k},\omega
\right) }{\omega +i\delta -\omega }.
\end{equation}%
We take for the spectral weight the Lorentzian form%
\begin{equation}
A_{\chi ,s}\left( \mathbf{k},\omega \right) =\frac{\Gamma /\pi }{\left(
\omega -\left( E_{\chi ,s}\left( \mathbf{k}\right) -\hslash v_{F}k_{F,\chi
}\right) /\hslash \right) ^{2}+\Gamma ^{2}},
\end{equation}%
where $\Gamma =1/2\tau $ and $\tau $ is the scattering time. After a lengthy
calculation, we recover the result (the real part of this conductivity is
derived in Ref. \onlinecite{Carbotte2016}).%
\begin{eqnarray}
\sigma _{\chi ,intra}\left( \omega \right) &=&\frac{e^{2}}{6\pi ^{2}\hslash
^{3}v_{F}}\frac{\tau }{1+\omega ^{2}\tau ^{2}}\left[ e_{F,\chi }^{2}+\frac{1%
}{3}\hslash ^{2}\omega ^{2}\right. \\
&&\left. +\frac{\hslash ^{2}}{4\tau ^{2}}+i\omega \tau \left( e_{F,\chi
}^{2}-e_{C}^{2}\right) \right] ,  \notag
\end{eqnarray}%
where $e_{C}$ is an energy cutoff in the valence band. The constant $%
e_{C}^{2}$ can be fixed by requiring that $\sigma _{\chi }\left( \omega
\right) $ satisfies the Kramers-Kroning relations. We find in this way that $%
e_{C}^{2}=\hslash ^{2}/12\tau ^{2}.$

\section{CONDUCTIVITY TENSOR OF THE WSM FOR $\mathbf{B}\neq 0$\ }

Formulas for the absorptive parts of the dynamic conductivity tensor due to
interband transitions in a simple two-node model with $b_{0}=0$ are given in
Ref. \onlinecite{Carbotte2013}. In this Appendix, we extend these results to
compute the complete dynamic conductivity tensor in the ultra quantum limit,
allowing for different populations for the two nodes. We use the
dimensionless variables defined in Eq. (\ref{reduce}) and the
current-current response tensor defined in Eq. (\ref{chid}). In the
ultra-quantum limit where the Fermi level is in the chiral level i.e.,%
\textit{\ }we must respect the condition $\left\vert e_{F,\chi
,r}\right\vert <\sqrt{2}.$ At $T=0$ K, we have for the filling factors

\begin{eqnarray}
\left\langle n_{0,\chi }\left( k\right) \right\rangle &=&\Theta \left(
k_{F,\chi }+\chi k\right) ,  \label{occup1} \\
\left\langle n_{n>0,s=-1,\chi }\left( k\right) \right\rangle &=&1,\,
\label{occup2} \\
\left\langle n_{n>0,s=1,\chi }\left( k\right) \right\rangle &=&0,
\label{occup3}
\end{eqnarray}%
where $\Theta \left( x\right) $ the step function. We have defined $%
k_{F,\chi }$ as the Fermi wave vector measured with respect to the Dirac
point in the node considered. Depending on the choice of $b_{0}$ and $%
e_{F,\chi },$ electrons or holes can be present in a node. After a lengthy
but straightforward calculation, we obtain the following expressions for the
interband contribution to the conductivity tensor and for $\omega >0$ (the
solution for $\omega <0$ can be obtained by symmetry)

\begin{widetext}

\begin{eqnarray}
\func{Re}\left[ \sigma _{xx}\left( \omega _{r}\right) \right]  &=&\frac{e^{2}%
}{\hslash \ell }\frac{1}{8\pi \omega _{r}}\left[ \Theta \left( \omega
_{r}-\left( x_{F}+\sqrt{x_{F}^{2}+2}\right) \right) +\Theta \left( \omega
_{r}-\left( -x_{F}+\sqrt{x_{F}^{2}+2}\right) \right) \right]  \\
&&+\frac{e^{2}}{\hslash \ell }\sum_{n>1}\frac{\omega _{r}^{4}-4}{16\pi
\omega _{r}^{3}}\sum_{\pm }\frac{\left( 1+\frac{x_{\pm }}{e_{n}\left( x_{\pm
}\right) }\right) \sqrt{1+\frac{x_{\pm }}{e_{n+1}\left( x_{\pm }\right) }}%
\sqrt{1+\frac{x_{\pm }}{e_{n-1}\left( x_{\pm }\right) }}}{\sqrt{\left(
\omega _{r}^{2}+2\right) ^{2}-8\omega _{r}^{2}n}}\Theta \left( \omega
_{r}-\Delta _{n}\left( 0\right) \right) ,  \notag
\end{eqnarray}%
\begin{equation}
\func{Im}\left[ \sigma _{xx}\left( \omega _{r}\right) \right] =\frac{e^{2}}{%
\hslash \ell }\frac{1}{8\pi ^{2}\omega _{r}}\left[ \log \left[ \left\vert 
\frac{\omega _{r}^{4}-4\omega _{r}^{2}x_{F}^{2}-4\omega _{r}^{2}+4}{4}%
\right\vert \right] +\omega _{r}^{2}\sum_{n>1}2\int_{-\infty }^{+\infty }dx%
\frac{\Lambda _{n}\left( x\right) }{\Delta _{n}\left( x\right) }\frac{1}{%
\omega _{r}^{2}-\Delta _{n}\left( x\right) ^{2}}\right] ,
\end{equation}%
\begin{equation}
\func{Re}\left[ \sigma _{xy}\left( \omega _{r}\right) \right] =-\frac{e^{2}}{%
\hslash \ell }\frac{1}{8\pi ^{2}\omega _{r}}\log \left[ \left\vert \frac{%
2-\omega _{r}^{2}+2\omega _{r}x_{F}}{2-\omega _{r}^{2}-2\omega _{r}x_{F}}%
\right\vert \right] ,  \label{sig1}
\end{equation}%
\begin{equation}
\func{Im}\left[ \sigma _{xy}\left( \omega _{r}\right) \right] =\frac{e^{2}}{%
\hslash \ell }\frac{1}{8\pi \omega _{r}}\left[ \Theta \left( \omega
_{r}-\left( x_{F}+\sqrt{x_{F}^{2}+2}\right) \right) -\Theta \left( \omega
_{r}-\left( -x_{F}+\sqrt{x_{F}^{2}+2}\right) \right) \right] ,  \label{sig2}
\end{equation}%
\begin{equation}
\func{Re}\left[ \sigma _{zz}\left( \omega _{r}\right) \right] =\frac{e^{2}}{%
\hslash \ell }\frac{1}{2\pi }\sum_{n>0}\frac{2n}{\omega _{r}^{2}\sqrt{\omega
_{r}^{2}-8n}}\theta \left( \omega _{r}-2\sqrt{2n}\right) ,
\end{equation}%
\begin{equation}
\func{Im}\left[ \sigma _{zz}\left( \omega _{r}\right) \right] =\frac{e^{2}}{%
\hslash \ell }\frac{\omega _{r}}{\pi ^{2}}\sum_{n>0}n\int_{0}^{+\infty }dx%
\frac{1}{\left( x^{2}+2n\right) ^{3/2}}\frac{1}{\omega _{r}^{2}-4\left(
x^{2}+2n\right) },
\end{equation}

\end{widetext}

These formulas are for one node and $x_{F}\equiv k_{F}\ell .$ We have also
defined the functions

\begin{equation}
e_{n}\left( x\right) =\sqrt{x^{2}+2n},\,
\end{equation}%
\begin{equation}
\,\Delta _{n}\left( x\right) =e_{n}\left( x\right) +e_{n-1}\left( x\right) ,
\end{equation}%
\begin{equation}
\,x_{n,\pm }=\pm \sqrt{\left( \frac{\omega _{r}^{2}+2}{2\omega _{r}}\right)
^{2}-2n},
\end{equation}%
\begin{eqnarray}
\Lambda _{n}\left( x\right) &=&\left( 1+\frac{x}{e_{n}\left( x\right) }%
\right) \sqrt{1+\frac{x}{e_{n+1}\left( x\right) }} \\
&&\times \sqrt{1+\frac{x}{e_{n-1}\left( x\right) }}.
\end{eqnarray}%
Using the definitions of the eigenspinors given in the main text, it is easy
to show that the only difference between the contribution of both nodes to
the conductivity comes from the difference in the local Fermi wave vector $%
k_{F,\chi }$ which depends on the Fermi level $e_{F}$ and $b_{0}$ through
Eq. (\ref{quatre}).

Defining a dimensionless conductivity tensor $\overline{\sigma }_{\chi
,\left( \alpha \beta \right) }$ by 
\begin{equation}
\sigma _{\chi ,\left( \alpha \beta \right) }\left( \omega \right) =\frac{%
e^{2}}{\hslash \ell }\overline{\sigma }_{\chi }^{\left( \alpha ,\beta
\right) }\left( \omega \right) ,
\end{equation}%
we have for the dielectric tensor of the two nodes%
\begin{equation}
\varepsilon _{\alpha \beta }=\delta _{\alpha \beta }+\frac{2\pi ^{2}\kappa i%
}{\omega _{r}v_{F,r}}\sum_{\chi }\overline{\sigma }_{\chi ,\left( \alpha
\beta \right) }\left( \omega \right) .  \label{sigma}
\end{equation}

As in Appendix A, the contribution of the intra-Landau-level transitions in $%
n=0$ to the conductivity tensor must be calculated by including disorder. In
our case, the intraband transitions are present only in the matrix element $%
\Gamma _{\chi ;0;0}^{\left( z\right) }\left( k\right) $ so that we need to
include disorder only in $\chi _{\chi }^{\left( z,z\right) }\left( \omega
\right) .$ We have for the intra-Landau-level current response function for
one node%
\begin{eqnarray}
\chi _{\chi ,\left( zz\right) }\left( \omega \right) &=&\frac{e^{2}}{2\pi
\ell ^{2}\hslash ^{2}}\int \frac{dk}{2\pi }\left\vert \Gamma _{\chi
;0,0}^{\left( z\right) }\left( k\right) \right\vert ^{2} \\
&&\times \int_{-\infty }^{+\infty }d\omega ^{\prime }\int_{-\infty
}^{+\infty }d\omega ^{\prime \prime }A_{\chi ,0}\left( k,\omega ^{\prime
}\right) A_{\chi ,0}\left( k,\omega ^{\prime \prime }\right)  \notag \\
&&\times \frac{F\left( \omega ^{\prime \prime }\right) -F\left( \omega
^{\prime }\right) }{\omega +i\delta +\omega ^{\prime \prime }-\omega
^{\prime }}.  \notag
\end{eqnarray}%
We again choose a Lorentzian for the spectral weight which is given by

\begin{equation}
A_{\chi ,0}\left( k,\omega \right) =\frac{\Gamma /\pi }{\left( \omega
-\left( E_{\chi ,0}\left( k\right) -e_{F,\chi }\right) /\hslash \right)
^{2}+\Gamma ^{2}}
\end{equation}%
and $e_{F,\chi }$ is the local Fermi level measured with respect to the
Dirac point in node $\chi $. At $T=0$ K, 
\begin{eqnarray}
\chi _{\chi ,\left( zz\right) }\left( \omega \right) &=&\frac{e^{2}}{\hslash
^{2}}\frac{1}{2\pi \ell ^{2}}\int \frac{dk}{2\pi }\left\vert \Gamma _{\chi
;0,0}^{\left( z\right) }\left( k\right) \right\vert ^{2} \\
&&\times \left[ \int_{-\infty }^{e_{F,\chi }/\hslash }d\omega ^{\prime
}a_{\chi }\left( k,\omega ^{\prime }\right) g_{\chi }\left( k,\omega
^{\prime }+\omega \right) \right.  \notag \\
&&+\int_{-\infty }^{e_{F,\chi }/\hslash }d\omega ^{\prime }a_{\chi }\left(
k,\omega ^{\prime }\right) g_{\chi }\left( k,\omega ^{\prime }-\omega \right)
\notag \\
&&-i\pi \left. \int_{e_{F}/\hslash -\omega }^{e_{F}/\hslash }d\omega
^{\prime }a_{\chi }\left( k,\omega ^{\prime }\right) a_{\chi }\left(
k,\omega ^{\prime }+\omega \right) \right] ,  \notag
\end{eqnarray}%
with the functions%
\begin{equation}
a_{\chi }\left( k,\omega \right) =\frac{\Gamma /\pi }{\left( \omega -E_{\chi
,0}\left( k\right) \left( k\right) /\hslash \right) ^{2}+\Gamma ^{2}}
\end{equation}%
and%
\begin{equation}
g_{\chi }\left( k,\omega \right) =\frac{\omega -E_{\chi ,0}\left( k\right)
/\hslash }{\left( \omega -E_{\chi ,0}\left( k\right) /\hslash \right)
^{2}+\Gamma ^{2}}.
\end{equation}

Evaluating these expressions analytically, we get for the intra-Landau-level
conductivity%
\begin{equation}
\sigma _{\chi ,\left( zz\right) }\left( \omega \right) =\frac{e^{2}}{\hslash
\ell }\frac{v_{F}\tau }{4\pi ^{2}\ell }\frac{1}{1-i\omega \tau }.
\end{equation}%
This expression is independent of $e_{F}$ and $b_{0}$ i.e., the same for
both nodes.


\begin{thebibliography}{99}
\bibitem{Review} For a review of Weyl semimetals, see, for example : P.
Hosur and X.-L. Qi, C. R. Physique \textbf{14}, 857-870 (2013); N. P.
Armitage, E. J. Mele, A. Vishwanath, Rev. Mod. Physics \textbf{90}, 15001
(2018).

\bibitem{Nielsen} H. B. Nielsen and M. Ninomiya, Phys. Lett. B\textbf{105},
219 (1981).

\bibitem{AHE} K.Y. Yang, Y.M. Lu, Y. Ran, Phys. Rev. B \textbf{84}, 075129
(2011); G. Xu, H. Weng, Z. Wang, X. Dai, Z. Fang, Phys. Rev. Lett. \textbf{%
107}, 186806 (2011); P. Goswami, S. Tewari, Phys. Rev. B \textbf{88}, 245107
(2013); A.A. Burkov, L. Balents, Phys. Rev. Lett. \textbf{107}, 127205
(2011); A.A. Zyuzin, S.Wu, A.A. Burkov, Phys. Rev. B \textbf{85}, 165110
(2012).

\bibitem{CME} J.H. Zhou, H. Jiang, Q. Niu, J.R. Shi, Chinese Phys. Lett. 30,
027101 (2013); Y. Chen, S. Wu, A.A. Burkov, Phys. Rev. B \textbf{88}, 125105
(2013).

\bibitem{FermiArc} X. Wan, A.M. Turner, A. Vishwanath, S.Y. Savrasov, Phys.
Rev. B 83, 205101 (2011); P. Hosur, Phys. Rev. B \textbf{86}, 195102 (2012).

\bibitem{ChiralAnomaly} H. Z. Lu, S. B. Zhang and S. Q. Shen, Phys. Rev. B 
\textbf{92}, 045203 (2015); F. Wilczek, Phys. Rev. Lett. \textbf{58}, 1799
(1987).

\bibitem{Spivak} D. T. Son and B. Z. Spivak, Phys. Rev. B \textbf{88},
104412 (2013).

\bibitem{Cheskis} D. Cheskis, Symmetry, \textbf{12}, 1412 (2020).

\bibitem{Han2022} X. Han, A. Markou, J. Stensberg, Y. Sun, C. Felser, and L.
Wu, Phys. Rev. B \textbf{105}, 174406 (2022).

\bibitem{Wilczek} F. Wilczek, Phys. Rev. Lett. \textbf{58}, 1799 (1987).

\bibitem{Wu} L. Wu, M. Salehi, N. Koirala, J. Moon, S. Oh, and N. P.
Armitage, Science \textbf{354}, 1124 (2016).

\bibitem{Burkov2012} A. A. Zyuzin and A. A. Burkov, Phys. Rev. B \textbf{86}%
, 115133 (2012); Y Chen, Si Wu, and A. A. Burkov, Phys. Rev. B \textbf{88},
125105 (2013).

\bibitem{Kargarian} M. Kargarian, M. Randeria and N. Trivedi, Sci. Rep. 
\textbf{5}, 12683 (2015).

\bibitem{Sonowal2019} K. Sonowal, A. Singh, and A. Agarwal, Phys. Rev. 
\textbf{100}, 085436 (2019).

\bibitem{Parent2020} J.-M. Parent, R. C\^{o}t\'{e}, and I. Garate, Phys.
Rev. B \textbf{102}, 245126 (2020).

\bibitem{Bertrand2019} S. Bertrand, J.-M. Parent, R. C\^{o}t\'{e}, and I.
Garate, Phys. Rev. B \textbf{100}, 075107 (2019).

\bibitem{Balkanski} M. Balkanski and R. F. Wallis, \textit{Many-Body Aspects
of Solid State Spectroscopy} (North-Holland, Amsterdam, 1986).

\bibitem{Carbotte2014} P. E. C. Ashby and J. P. Carbotte, Phys. Rev. B 
\textbf{89}, 245121 (2014).

\bibitem{Carbotte2016} C. J. Tabert, J. P. Carbotte, and E. J. Nicol, Phys.
Rev. B \textbf{93}, 085426 (2016).

\bibitem{Carbotte2018} S. P. Mukherjee and J. P. Carbotte, Phys. Rev. B 
\textbf{97}, 035144 (2018).

\bibitem{Carbotte2021} A. Singh and J. P. Carbotte, Phys. Rev. B \textbf{103}%
, 075114 (2021).

\bibitem{Goerbig2016} S. Tchoumakov, M. Civelli and M. O. Goerbig, Phys.
Rev. Lett. \textbf{117}, 086402 (2016).

\bibitem{MaPesin2015} J. Ma and D. A. Pesin, Phys. Rev. B \textbf{92},
235205 (2015).

\bibitem{Zhong2016} S. Zhong, J. E. Moore, and I. Souza, Phys. Rev. Lett. 
\textbf{116}, 077201 (2016).

\bibitem{Kotov2016} O. V. Kotov and Y. E. Lozovik, Phys. Rev. B \textbf{93},
235417 (2016).

\bibitem{Sa2021} D. Sa, Eur. Phys. J. B \textbf{94}, 31 (2021).

\bibitem{Deng2021} K. Deng, J. S. Van Dyke, D. Minic, J. J. Heremans, and E.
Barnes, Phys. Rev. B \textbf{104}, 075202 (2021).

\bibitem{BornWolf} M. Born and W. Wolf, \textit{Principles of optics},
(Cambridge University Press, 1980).

\bibitem{Levy2020} A. L. Levy, A. B. Sushkov, F. Liu, B. Shen, N. Ni, H. D.
Drew and G. S. Jenkins, Phys. Rev. B \textbf{101}, 125102 (2020).

\bibitem{Arnold2016} F. Arnold, M. Naumann, S.-C. Wu, Y. Sun, M. Schmidt, H.
Borrmann, C. Felser, B. Yan, and E. Hassinger, Phys. Rev. Lett. \textbf{117}%
, 146401 (2016).

\bibitem{Plie2017} P. Li, Y. Wen, X. He, Q. Zhang, C. Xia, Z.-M. Yu, S. A.
Yang, Z. Zhu, H. N. Alshareef, and X.-X. Zhang, Nat. Commun. \textbf{8},
2150 (2017).

\bibitem{Huang2021} X. Huang, H. Geng, and L. Sheng, Phys. Rev. B \textbf{103%
}, 115208 (2021).

\bibitem{PlasmonBnul} S. Das Sarma, E. H. Hwang, Phys. Rev. Lett. \textbf{102%
}, 206412 (2009); M. Lv, S. C. Zhang, Int. J. Mod. Phys. B \textbf{27},
1350177 (2013) ; J. Hofmann and S. Das Sarma, Phys. Rev. B \textbf{91},
241108(R) (2015).

\bibitem{Zhou2015} J. Zhou, Hao-Ran Chang, D. Xiao, Phys. Rev. B \textbf{91}%
, 035114 (2015).

\bibitem{Helicon2015} F. M. D. Pellegrino, M. I. Katsnelson, and M. Polini,
Phys. Rev. B \textbf{92}, 201407(R) (2015).

\bibitem{Zebin2017} Z. Qiu, G. Cao, and X.-G. Huang, Phys. Rev. D \textbf{95}%
, 036002 (2017).

\bibitem{Okamura2020} Y. Okamura, S. Minami, Y. Kato, Y. Fujishiro, Y.
Kaneko, J. Ikeda, J. Muramoto, R. Kaneko, K. Ueda, V. Kocsis, N. Kanazawa,
Y. Taguchi, T. Koretsune, K. Fujiwara, A. Tsukazaki, R. Arita, Y. Tokura and
Y. Takahashi, Nat. Commun. \textbf{11}, 4619 (2020).

\bibitem{Wang2017} Q. Wang, Y. Xu, R. Lou, Z. Liu, M. Li, Y. Huang, D. Shen,
H. Weng, S. Wang, and H. Lei, Nat Commun \textbf{9}, 3681 (2018).

\bibitem{Wang2019} L.-L. Wang, N. H. Jo, B. Kuthanazhi,Y. Wu, R. J.
McQueeney, A. Kaminski, and P. C. Canfield, Phys. Rev. B \textbf{99}, 245147
(2019).

\bibitem{Soh2019} %
Soh,J.R.,deJuan,F.,Vergniory,M.G.,Schroter,N.B.M.,Rahn,M.C.,
Yan,D.Y.,Jiang,J.,Bristow,M.,Reiss,P.,Blandy,J.N.,Guo,Y.F.,Shi,Y.G.,Kim,T.K., McCollam,A.,Simon,S.H.,Chen,Y.,Coldea,A.I.,Boothroyd,A.T., Phys. Rev. B 
\textbf{100}, 201102(R), (2019).

\bibitem{Nie2022} S. Nie, T. Hashimoto, and F. B. Prinz, Phys. Rev. Lett. 
\textbf{128}, 176401 (2022).

\bibitem{Carbotte2013} P. E. C. Ashby and J. P. Carbotte, Phys. Rev. \textbf{%
87}, 245131 (2013).
\end{thebibliography}
\end{document}